\PassOptionsToPackage{dvipsnames,table}{xcolor}

\documentclass[sigconf]{acmart}

\copyrightyear{2025}
\acmYear{2025}
\setcopyright{cc}
\setcctype{by}
\acmConference[CCS '25]{Proceedings of the 2025 ACM SIGSAC Conference on Computer and Communications Security}{October 13--17, 2025}{Taipei, Taiwan}
\acmBooktitle{Proceedings of the 2025 ACM SIGSAC Conference on Computer and Communications Security (CCS '25), October 13--17, 2025, Taipei, Taiwan}
\acmISBN{979-8-4007-1525-9/2025/10}
\acmDOI{10.1145/3719027.3765136}

\usepackage{hyperref}

\usepackage{tikz}
\usepackage{amsmath}
\usepackage{mdframed}
\usepackage{pifont}

\usepackage{enumitem}
\usepackage{longtable}
\usepackage{caption}
\usepackage{listings}
\usepackage{tabularx}
\usepackage{ragged2e}

\usepackage[capitalise]{cleveref}
\usepackage{multicol}
\usepackage{multirow}

\usepackage{algorithm}
\usepackage{algpseudocode}

\setlist[enumerate]{noitemsep,topsep=2pt,leftmargin=*}
\setlist[itemize]{noitemsep,topsep=2pt,leftmargin=*}

\newcolumntype{L}[1]{>{\RaggedRight\arraybackslash}p{#1}} 
\newcolumntype{C}[1]{>{\Centering\arraybackslash}p{#1}}    
\newcolumntype{R}[1]{>{\RaggedLeft\arraybackslash}p{#1}}   

\lstdefinelanguage{js}{
	keywords={typeof, new, true, false, try, catch, function, return, null, undefined, catch, switch, var, if, in, for, while, do, else, case, break,let, const, throw, with, await, delete},
	keywordstyle=\color{Maroon}\bfseries,
	ndkeywords={class, export, boolean, throw, implements, this, async},
	ndkeywordstyle=\color{darkgray}\bfseries,
	identifierstyle=\color{black},
	sensitive=false,
	comment=[l]{//},
	morecomment=[s]{/*}{*/},
	commentstyle=\color{darkgray}\ttfamily,
	stringstyle=\color{OliveGreen}\ttfamily,
	escapeinside={/*\#}{\#*/},
	morestring=[b]',
	morestring=[b]",
	morestring=[b]`
}
\newcommand{\tightpar}[1]{{\smallskip \noindent\bf #1}}

\newcommand{\new}[1]{\textcolor{blue}{#1}}
\newcommand{\newer}[1]{\textcolor{cyan}{#1}}
\newcommand{\changed}[1]{\textcolor{red}{#1}}
\renewcommand{\new}[1]{#1}
\renewcommand{\newer}[1]{#1}
\renewcommand{\changed}[1]{#1}

\newcommand{\apparmor}{AppArmor}
\newcommand{\ebpf}{eBPF}
\newcommand{\esmodules}{\textsc{ESModules}}
\newcommand{\commonjs}{\textsc{CommonJS}}
\newcommand{\landlock}{Landlock LSM}
\newcommand{\npm}{npm}
\newcommand{\pypi}{PyPI}
\newcommand{\rubygems}{RubyGems}
\newcommand{\seccomp}{seccomp}
\newcommand{\yarn}{yarn}

\newcommand*\emptycirc[1][1ex]{%
  \raisebox{-0.2\height}{\tikz\draw (0,0) circle (#1);}}
\newcommand*\halfcirc[1][1ex]{%
  \raisebox{-0.2\height}{%
    \begin{tikzpicture}
      \draw[fill] (0,0)-- (90:#1) arc (90:270:#1) -- cycle ;
      \draw (0,0) circle (#1);
    \end{tikzpicture}}}
\newcommand*\fullcirc[1][1ex]{%
  \raisebox{-0.2\height}{\tikz\fill (0,0) circle (#1);}}

\newcommand{\xmark}{\ding{55}}

\begin{document}

\date{}

\title{NodeShield: Runtime Enforcement of Security-Enhanced SBOMs for Node.js}

\author{Eric Cornelissen}
\email{ericco@kth.se}
\affiliation{%
  \institution{KTH Royal Institute of Technology}
  \city{Stockholm}
  \country{Sweden}
}
\author{Musard Balliu}
\email{musard@kth.se}
\affiliation{%
  \institution{KTH Royal Institute of Technology}
  \city{Stockholm}
  \country{Sweden}
}
\renewcommand{\shortauthors}{Eric Cornelissen and Musard Balliu}

\begin{abstract}
The software supply chain is an increasingly common attack vector for malicious actors. The Node.js ecosystem has been subject to a wide array of attacks, likely due to its size and prevalence. To counter such attacks, the research community and practitioners have proposed a range of static and dynamic mechanisms, including process- and language-level sandboxing, permission systems, and taint tracking. Drawing on valuable insight from these works, this paper studies a runtime protection mechanism for (the supply chain of) Node.js applications with the ambitious goals of compatibility, automation, minimal overhead, and policy conciseness.

Specifically, we design, implement and evaluate NodeShield, a protection mechanism for Node.js that enforces an application's dependency hierarchy and controls access to system resources at runtime. We leverage the up-and-coming SBOM standard as the source of truth for the dependency hierarchy of the application, thus preventing components from stealthily abusing undeclared components. We propose to enhance the SBOM with a notion of capabilities that represents a set of related system resources a component may access. Our proposed SBOM extension, the Capability Bill of Materials or CBOM, records the required capabilities of each component, providing valuable insight into the potential privileged behavior. NodeShield enforces the SBOM and CBOM at runtime via code \emph{outlining} (as opposed to inlining) with no modifications to the original code or Node.js runtime, thus preventing unexpected, potentially malicious behavior. Our evaluation shows that NodeShield can prevent over 98\% out of 67 known supply chain attacks while incurring minimal overhead on servers at less than 1ms per request. We achieve this while maintaining broad compatibility with vanilla Node.js and a concise policy language that consists of at most 7 entries per dependency.
\end{abstract}




\maketitle

\section{Introduction}
\label{section:introduction}

Prominent attacks such as SolarWinds, Log4Shell, and XZ Utils have brought software supply chain security to a top priority for both the academic community and industry practitioners. A recent approach to help mitigate security risks has been the introduction of the Software Bill of Materials (SBOM), a machine-readable inventory of the ingredients of a software application, including the components, their metadata, and their internal relationships. While the effectiveness of SBOMs is still subject to debate~\cite{Okhravi25}, the community agrees that they increase software transparency, contributing as a reactive measure to identify one-day vulnerabilities, e.g., in applications using vulnerable components. This paper envisions an enhancement of SBOMs with capabilities (a list of sensitive APIs that components use), dubbed CBOM, and takes it a step further by enforcing these capabilities at runtime, thus making CBOM and SBOM a proactive measure against supply chain attacks.

Node.js and the \npm~ecosystem of software components, or packages, have been a particularly attractive target for supply chain attacks~\cite{duan2020maloss,ohm2020backstabber,ladisa2023sok}. Beyond the security challenges of JavaScript, this is facilitated by a culture encouraging the use of many packages and automatic dependency updates. Research shows that a single package trusts on average 79 other third-party packages published by 40 maintainers~\cite{Zimmermann19}. This creates an invisible attack surface, where a single compromise has widespread consequences~\cite{eventstreamincident,purescript,eslint,snykelectron,getcookies}. To further exacerbate the risks, Node.js applications typically run with full access to system resources, which third-party packages inherit. In this context, a security system specifically designed for the Node.js supply chain is crucial. By enforcing the SBOM hierarchy and the CBOM capabilities on package behavior---such as restricting access to files, network, or processes---we can significantly reduce the risk of malicious updates while enhancing transparency and control within the ecosystem.

NodeShield~\cite{artifact} contributes with the design and implementation of a practical system that focuses on the supply chain of Node.js applications while ensuring compatibility, automation, minimal overhead, policy conciseness, and robustness to attacks. To the best of our knowledge, no system meets all these goals. \changed{As shown in our evaluation (\Cref{section:evaluation}), related works on lightweight permission systems for Node.js~\cite{ferreira2021containing,ohm2023you} show limitations with regard to compatibility, policies, and attack robustness.} Finer-grained approaches, e.g. language-level sandboxing~\cite{de2014nodesentry,vasilakis2018breakapp,vasilakis2021preventing,ahmadpanah2021sandtrap,lavamoat} or taint analysis~\cite{SchoepeBPS16,augur,jalangi,pmforce,ShcherbakovMB24,cassel2023nodemedic}, offer stronger protection at the expense of performance, false positives, automation, and policy conciseness.

At the heart of NodeShield lies a non-intrusive package-level instrumentation, called outlining, that enforces the SBOM and CBOM policies at runtime, with no modifications to the original code or Node.js runtime. We achieve this by a principled design and security analysis that considers access to system resources and enforcement bypasses. In particular, we develop a novel lexical scoping-based approach that reduces the impact of bypassing the enforcement and provides support for both legacy and modern Node.js modules, \commonjs~and \esmodules. Through a methodological analysis of Node.js APIs and related systems, we propose seven security-sensitive capabilities for the CBOM, thus creating a concise policy language. NodeShield enforces these capabilities through a comprehensive coverage of three avenues: built-in modules, global variables, and native bindings. Moreover, it supports automated policy generation along with features that facilitate CBOM presentation at different granularity levels to aid adoption for developers.

To evaluate the effectiveness of NodeShield, we contribute with a benchmark of 67 known supply chain attacks and show that it can prevent 98.51\%. We further find that NodeShield can reduce the impact of arbitrary code execution vulnerabilities by detecting 87.50\% of exploits from SecBench.js~\cite{bhuiyan2023secbench}. Our performance evaluation on long-lived applications shows a response time overhead below 1ms and a throughput reduction of up to 360 requests per second. We also find that NodeShield is broadly compatible with software written for Node.js, despite some incompatible coding patterns being used in practice. We evaluate the maintenance effort of CBOMs to find that application developers need to review less than 1 capability per dependency update on average, and between 0 and 13 capabilities for new dependencies. Our robustness analysis against sandbox bypasses shows that NodeShield prevents all attacks within our threat model. Finally, we perform an end-to-end analysis of the attack on Copay~\cite{eventstreamincident}, showing the usefulness of NodeShield in practice. In sum, we offer the following contributions:

\begin{itemize}
  \item The first Node.js runtime protection tool that comprehensively protects both \commonjs~and \esmodules-based source code.
  \item A novel technique to contain Node.js sandbox breakouts, providing better guarantees than related work on malware protection.
  \item An SBOM extension to capture the privileges required by supply chains components at the granularity of packages.
  \item A comprehensive evaluation of effectiveness, performance, compatibility, maintainability, and robustness to attacks.
\end{itemize}

\noindent NodeShield and all experiments are available at \url{https://github.com/KTH-LangSec/nodeshield}.

\section{Preliminaries}
\label{section:preliminaries}

\tightpar{Node.js}
Node.js is a JavaScript runtime intended for server-side applications, built on the V8 engine from the Chromium browser extended with APIs for access to system resources. Node.js supports two module systems: \commonjs~and \esmodules. The former leveraging legacy JavaScript features to establish namespaces and supports importing modules using the \verb|require| (or \verb|import|) function. The latter is a new format with native namespacing and syntax for importing and exporting modules (besides the \verb|import| function).

Node.js applications rely on third-party dependencies, or \emph{packages}. Packages are usually distributed via a registry (e.g., \verb|npmjs.com|) and managed by a package manager (e.g., \npm~or \yarn). Direct dependencies of an application are stored in the manifest file with, optionally, all dependencies listed in a lockfile (\verb|package.json| and \verb|package-lock.json| resp. for \npm). Node.js uses the \verb|node_modules| directory to store and access third-party packages. Packages can be reused by name through the \verb|require| function (\verb|require("vm")| in \commonjs), \emph{import syntax} (\verb|import vm from "vm"| in \esmodules), or \verb|import| function (\verb|import("vm")| in both).

\tightpar{System interface}
Node.js provides access to system resources in two avenues. First, the Node.js built-in modules provide APIs to interact with the underlying system (e.g., the \verb|fs| module). These can be imported like third-party packages. Second, Node.js extends the list of global variables with built-in objects, APIs, and properties that provide system access (e.g., the \verb|process| global).
Either way, this interface is enabled by bindings between the JavaScript world and underlying C++ codebase, which are accessible directly through the (undocumented) \verb|process.binding| function.

\tightpar{Code evaluation}
There are two main APIs for dynamic code evaluation in Node.js. First, the \verb|eval| function (and friends, like \verb|new Function|) allow for basic dynamic code evaluation~\cite{tc39}. A call to \verb|eval| evaluates a string as JavaScript in the current lexical scope. Use of \verb|require| or \verb|import| with \verb|eval| is possible, but the \esmodules~import and export syntax is not.

Secondly, Node.js provides a built-in module for code evaluation called \verb|vm|. This module offers an API for evaluating both \commonjs~and \esmodules~JavaScript in a new V8 context while allowing object sharing between contexts. A V8 context is an separate interpreter context within an interpreter process, allowing multiple contexts to share JavaScript objects within the same process. Each V8 context, including the top-level context created when a Node.js application starts, can be configured independently.

\tightpar{SBOM}
A Software Bill Of Materials (SBOM) is a document that lists all the components used in a software artifact. For each component it may include various pieces of information ranging from checksums to licensing information. Additionally, a good SBOM includes the dependency hierarchy of all components, specifying which components each one uses. There exist two widely-used specifications for SBOMs, SPDX and CycloneDX, along with myriad of automated tooling to generate SBOMs for software artifacts.

\section{Overview}
\label{section:overview}

This section provides a high-level overview of the research problem and solution. It also discusses the threat model and the proposed usage of capabilities as a security enhancement of SBOMs.

\subsection{Challenges and Solution Overview}
\label{section:challenges}

\tightpar{The problem}
We use the example of the \verb|rate-map| package to illustrate our research problem. This is a very simple benign package that maps number in the range of \verb|[0,1]| to a new value within a given range. It is written using a combination of \commonjs~and \esmodules. The benign version v1.0.2 of \verb|rate-map| depends on one other package, \verb|append-type|, and implements a single function in JavaScript with no Node.js-specific functionality (line 4 in \Cref{listing:rate-map}).

\Cref{listing:rate-map} shows the compromised version v1.0.3 breaking the functionality of the \verb|purescript-installer| package which used \verb|rate-map| among its dependencies~\cite{snykratemap,purescript}. The malicious version adds an explicit dependency on the package \verb|terser| to its package manifest (not shown). However, it does not use \verb|terser|, instead it covertly resolves the package \verb|dl-tar| at runtime (line 1). This behavior hinders supply chain transparency and may be challenging to detect due to the use of obfuscation. Thus, our first challenge is to identify and enforce the expected dependency hierarchy of applications, and detect usage of packages outside this hierarchy.

Secondly, the malicious version introduces the use of Node.js's \verb|fs| module (line 2), which grants access to privileged file system APIs (line 3). Unfortunately, the current use of Node.js and \npm~(or similar) provides no insight into the use of privileged APIs. Thus, our second challenge is to identify and enforce the expected use of access to privileged resources on the supply chain of applications.

The attack combines these two malicious changes to modify the implementation of the \verb|dl-tar| package at runtime. The attacker's goal is to cause a denial of service, e.g., on the installation of the PureScript application. As highlighted, the attack can be detected at two stages: when an unexpected third-party dependency is resolved or when a new privileged API was adopted by the package.

\begin{figure}[t]
\centering
\begin{lstlisting}[caption={Simplified \texttt{rate-map} malware.},label={listing:rate-map}]
const px = require.resolve(Buffer.from([100, 108, 45, 116, 97, 114]/*=dl-tar*/).toString());
const fs = require("fs");
fs.writeFileSync(px, fs.readFileSync(px, "utf8").replace(a, b));
module.exports = function rateMap(val, start, end) {... return start + val * (end - start)}
\end{lstlisting}
\end{figure}

\begin{figure}[t]
  \centering
  \includegraphics[width=0.25\textwidth]{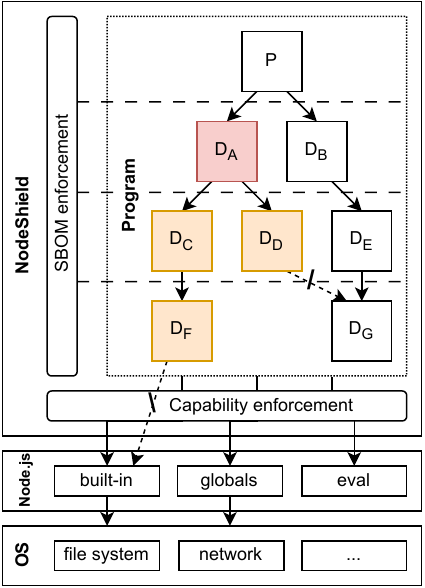}
  \caption{Overview and Threat Model of NodeShield.}
  \label{figure:high-level}
\end{figure}

\tightpar{Our solution}
To address this problem we aim to design a system that protects against software supply chain attacks as illustrated in \Cref{figure:high-level}. At a high level this is achieved by 1) enforcing the application dependency hierarchy according to its SBOM and 2) enforcing restrictions on the use of privileged APIs per dependency---modeled as ``capabilities''---covered in detail in \Cref{section:implementation:enforcement}. This is enhanced by a principled design and security analysis, and 3) applying hardening techniques to prevent bypassing our enforcement, as discussed in \Cref{section:implementation:security-analysis}, and a thorough empirical evaluation, cf. \Cref{section:evaluation}.

NodeShield builds on related work on runtime hardening for software supply chain attacks in JavaScript~\cite{vasilakis2021preventing,lavamoat,ferreira2021containing,ohm2023you} and language-level JavaScript sandboxing~\cite{ahmadpanah2021sandtrap,vasilakis2018breakapp}. In particular, we extend these techniques to support \esmodules~and propose a novel lexical scoping-based approach that reduces the impact of sandbox breakouts. The result is a system that hardens against supply chain attacks and bridges the existing gap with JavaScript sandboxes.

In practical terms, NodeShield enforces the dependency hierarchy by controlling what a package is allowed to import. This is achieved by trapping and validating all import attempts (covering \verb|require|, \verb|import|, and import syntax) on a per-dependency basis. For \verb|rate-map|, NodeShield would have prevented the covert import of \verb|dl-tar|, as represented by the line from $D_D$ to $D_G$ in \Cref{figure:high-level}.

The enforcement of capabilities is achieved through a comprehensive coverage of three separate avenues: built-in modules, global variables, and native bindings. First, the use of privileged built-in modules is controlled in the same way as the dependency hierarchy. Second, access to global variables is controlled by adapting the \verb|vm| module to create package-specific global namespaces. Lastly, access to bindings is controlled by overriding the \texttt{process.binding} API. In the case of \verb|rate-map|, NodeShield would have caught the newly introduced use of \verb|fs|. In \Cref{figure:high-level}, this is represented by the crossed-out line from $D_F$ to the Node.js built-in modules.

Finally, NodeShield is hardened using primordials~\cite{primordials}, \verb|null| prototypes, object freezing, and lexical scoping. This prevents application code from manipulating the policy and achieving anything by breaking out of its \verb|vm| context.

\subsection{Threat Model}
\label{section:overview:threat-model}

Our threat model focuses on server-side Node.js applications that use untrusted third-party dependencies. As illustrated by the highlighted squares in \Cref{figure:high-level}, the attacker controls an arbitrary set of packages in the dependency hierarchy of an application (which includes its transitive dependencies). We assume Node.js, the package manager (e.g., \npm~or \yarn), the package registry (e.g., \npm), the SBOM generator, and the supply chain of NodeShield are not compromised. \new{Since application dependencies may be openly malicious, developers are expected to review capabilities. This can be a significant upfront investment for pre-existing apps, but \Cref{section:evaluation:maintenance} shows it is manageable as a continuous task.}

The goal of the attacker, based on empirically-observed behavior of real JavaScript malware~\cite{ohm2020backstabber}, is to break the confidentiality or integrity of the system on which the application runs by using capabilities outside the set granted to attacker-controlled dependencies.

\begin{table*}
  \centering
  \rowcolors{2}{white}{gray!20}
  \begin{tabular}{ c | p{.40\textwidth} | p{.23\textwidth} | p{.20\textwidth} }
    \hline
    \textbf{Capability} & \textbf{Modules}                               & \textbf{Global Variables}                 & \textbf{Bindings}                    \\
    \hline
    addon               & \texttt{*.node}                                &                                           &                                      \\
    code                & \texttt{node:vm}                               & \texttt{eval}, \texttt{Function}          &                                      \\
    command             & \texttt{node:child\_process},
                          \texttt{node:worker\_threads}                  &                                          & \texttt{spawn}, \texttt{spawn\_sync}  \\
    crypto              & \texttt{node:crypto}                           & \texttt{Crypto}, \texttt{crypto},
                                                                           \texttt{CryptoKey}, \texttt{SubtleCrypto} & \texttt{crypto}                      \\
    file-system         & \texttt{node:fs}, \texttt{node:fs/promises}    &                                           &                                      \\
    network             & \texttt{node:net}, \texttt{node:http},
                          \texttt{node:http2}, \texttt{node:https},
                          \texttt{node:dns}, \texttt{node:/promises},
                          \texttt{node:tls}, \texttt{node:dgram}         & \texttt{fetch}                            &                                      \\
    system              & \texttt{node:os}, \texttt{node:process}        & \texttt{process}                          &                                      \\
    \hline
  \end{tabular}
  \caption{
    Capability mapping to modules, global variables, and binding.
    Built-in module names are listed as \texttt{node:*} but are also accessible without the \texttt{node:} prefix.
  }
  \label{table:capabilities}
\end{table*}

We consider the act of providing any data or reference to a dependency as entrusting that dependency tree with it, including capability-bearing references. The justification for this is that developers use dependencies to perform actions on the provided data using any capabilities granted.
We note that the global namespace of the application is considered as an intentional interface between components, and is therefore unprotected. As a result malicious code could exploit gadgets, e.g., prototype pollution~\cite{ShcherbakovBS23}, through this avenue. We further motivate this decision in \Cref{section:implementation:security-analysis}.

\subsection{Enhancing SBOM with Capabilities}
\label{section:overview:capabilities}

A key goal for NodeShield is to provide a simple and comprehensible policy language for developers, while effectively and efficiently preventing malware in the software supply chain. To this end, we argue that an enhancement of SBOMs with a Capability Bill of Materials (CBOM), at the granularity of packages, strikes the right balances between these requirements, as witnessed by our evaluation in \Cref{section:evaluation}. We use the term ``capability'' to refer to a group of related privileged operations.

We determine the set of capabilities through a comprehensive review of existing proposals across different ecosystems. Specifically, we use the following methodology: First, we collect a set of potential capabilities from related work in the area of supply chain security~\cite{ferreira2021containing,huang2024donapi} and related tooling (Node.js' and Deno's permission system~\cite{node-docs,doglio2020introducing}, Capslock~\cite{capslock}, and Cackle~\cite{cackle}). Then we filter this set of potential capabilities by omitting those unrelated to supply chain attacks according to the SAP supply chain attack tree~\cite{ladisa2023sok}, Socket.dev supply chain alerts~\cite{socketalerts}, and the OpenSSF metrics and data workgroup's OSE threat model~\cite{ossfose}.

From this analysis, we settle for a list of 7 capabilities listed below. \Cref{table:capabilities} provides an overview of the mapping from capability to modules, global variables, and bindings in Node.js.

\begin{itemize}
  \item The \texttt{addon} capability represents the ability of a dependency to load native code. This is included because native code can bypass our enforcement, enabling misuse of other capabilities.
  \item The \texttt{code} capability represents the ability to dynamically evaluate code. While dynamic code is subject to the other capabilities, this capability enables explicit control, thus facilitating manual review and defense in depth.
  \item The \texttt{command} capability represents the ability to run a command as subprocess. Subprocesses can bypass our enforcement, enabling the potential use of any other capability.
  \item The \texttt{crypto} capability controls usage of cryptography-related functionality. This is included because of obfuscation and ransomware attacks utilizing cryptography.
  \item The \texttt{file-system} capability represents all access to the file system. This is included because malicious packages may read sensitive data, e.g., cryptographic material, from the file system.
  \item The \texttt{network} capability represents all access to the network. This is included because malicious packages may exfiltrate sensitive data or fetch malicious code or commands from a remote server.
  \item The \texttt{system} capability represents all access to system and environment information. Malicious packages may read sensitive data, e.g., authentication tokens, from the environment variables.
\end{itemize}

\section{NodeShield}
\label{section:implementation}

This section details the design and implementation of NodeShield~\cite{artifact}. \Cref{figure:workflow} present an overview of NodeShield's architecture and workflow. NodeShield takes as input the source code of an application (including own and dependencies' source code), the application's SBOM (including the hierarchy of dependencies), and (optionally) the application's CBOM. From this it automatically creates a clone of the original project which supports runtime enforcement of the SBOM and CBOM. As an intermediate step, it creates a module-granular policy representation of the SBOM and CBOM, for both the application and every (transitive) dependency. If no CBOM is provided, NodeShield infers it statically---as described in \Cref{section:implementation:capability-inference}---before policy creation. The policy specifies what each module is allowed to import---files, third-party packages, built-in modules---what global variables it is allowed to access, and what bindings it is allowed to use.

To enforce the policy at runtime, NodeShield extends (but does not modify) the original source code on a per-file basis such that the file's code is evaluated in a \texttt{vm} context that enforces the policy of the module to which the file belongs. In contrast to inlining, we dub this process \emph{outlining} and describe it in \Cref{section:implementation:outlining}. The enforcement uses unmodified Node.js and \texttt{vm} APIs (\Cref{section:implementation:enforcement}) and is hardened to prevent policy manipulation and bypasses (\Cref{section:implementation:security-analysis}).

\subsection{Outlining}
\label{section:implementation:outlining}

To enforce the SBOM and CBOM at runtime, NodeShield outlines the original source code within enforcement code. The details of the enforcement are discussed in \Cref{section:implementation:enforcement}, here we present the transformation of the original source code to the \emph{outlined} one. As a basis, the outlining starts by creating a clone of the original application project. This clone only covers JavaScript source code files, JSON files, and native extensions. Other files in the project are accounted for by setting the working directory of the final program and rewriting paths present in Node.js APIs as part of the outlining. The cloned project can be run using vanilla Node.js.

The outlining procedure targets all JavaScript source code files, while JSON files and native extensions remain unchanged. In particular, the original source code is moved into a (multi-line) string that will be evaluated using \texttt{vm}. Prior to evaluation, NodeShield cleans up and prepares the host context (i.e., the top-level Node.js context) and the guest context (i.e., the \texttt{vm} context evaluating the original source code). \Cref{listing:outlining} illustrates the outlined code.

\begin{figure}[t]
\centering
\begin{lstlisting}[
  caption={Outlining Process of NodeShield.},
  label={listing:outlining}
]
INPUT: module, SBOM, CBOM
once(capture primordials)
once(capture and unbind sensitive global variables)
if (cjs) capture and unbind CJS global variables

Ia = files in module
Ib = packages according to SBOM
Ic = allowed-by-default built-in modules
Id = privileged built-in modules according to CBOM
I = [...Ia, ...Ib, ...Ic, ...Id]

Ba = allowed-by-default bindings
Bb = privileged bindings according to CBOM
B = [...Ba, ...Bb]
process.binding = (s) => if (s in B) process.binding(s) else halt

Ga = allowed-by-default global variables
Gb = privileged global variables according to CBOM
G = {...Ga, ...Gb}
if (cjs) {
  G.require = (s) => if (s in I) import(s) else halt
  ... other CJS variables
}

C = vm.NewContext(G, {
  import: (s) => if (s in I) import(s) else halt })
vm.Link((s) => if (s in I) import(s) else halt)
vm.Eval('<original source code>', C)
\end{lstlisting}
\end{figure}

First, before any untrusted code is evaluated, the cloned project obtains references to primordials (line 2, more on this in \Cref{section:implementation:security-analysis}) and privileged global variables (as defined in \Cref{table:capabilities}) which are also removed from the global namespace (line 3). Both need to happen only once. Next, for every \commonjs~file, references to the \commonjs~variables (\texttt{exports}, \texttt{module}, \texttt{require}, \texttt{\_\_dirname}, \texttt{\_\_filename}) must also be obtained and removed from the global namespace (line 4).

Then, it initializes an allowlist \texttt{I} for importing as the combination of 1) the JavaScript and JSON files in the module (line 6), 2) the packages that are listed as its direct dependencies in the SBOM (line 5), 3) the non-privileged built-in modules (line 8)---spanning all built-in modules not declared in \Cref{table:capabilities}---and 4) the allowed built-in modules in accordance with the CBOM, following the mapping of \Cref{table:capabilities} (line 9). The resulting list (line 10) is used when importing (line 26 and 27) or requiring (line 21) to determine if the specified identifier is allowed by the policy.

Next, the allowlist \texttt{B} of bindings is initialized as the combination of 1) the non-privileged bindings (line 12)---spanning all bindings not declared in \Cref{table:capabilities}---and 2) the allowed bindings in accordance with the CBOM, following the mapping of \Cref{table:capabilities} (line 13). The resulting list (line 14) is used when accessing bindings through the \texttt{process} object (line 15).

Next, the global namespace \texttt{G} for the guest context is initialized as the combination of 1) all non-privileged global variables (line 17)---spanning all global variables not declared in \Cref{table:capabilities}, 2) the allowed global variables in accordance with the CBOM, following the mapping of \Cref{table:capabilities} (line 18), and, for \commonjs, 3) the \commonjs-specific variables (line 21-22). The latter are not provided as-is, instead \texttt{require} is modified to check against the import allowlist and to present an empty module cache (which otherwise grants direct access to all previously loaded modules). Moreover, \texttt{module}, \texttt{\_\_dirname}, and \texttt{\_\_filename} are modified to reflect the original source code file paths.

Finally, the context for \texttt{vm} is initialized (line 25-26) and the source code evaluated with \texttt{vm} using this context (line 28). With the outlining process applied to all JavaScript files in the project, the policy enforcement is encoded directly into the program, which can now be executed safely as a regular Node.js application.

\tightpar{Example}
\new{For \Cref{listing:rate-map}, \Cref{listing:outlining} would concretely instantiate with the code on line 28; \texttt{Ia} as [index.js, package.json], \texttt{Ib} as [append-type, terser], \texttt{Ic} as [assert, console, \textit{etc.}], and \texttt{Id} as [] (line 6-9); \texttt{Bb} as [] (line 12-13); \texttt{Ga} as [atob, console, \textit{etc.}] and \texttt{Gb} as [] (line 17-18). These lists are the policy at runtime---derived directly from the SBOM and CBOM---e.g., requiring \texttt{fs} (line 2, \Cref{listing:rate-map}) is disallowed on line 21 because it is not in the allowlist \texttt{I}.}

\begin{figure}[t]
  \centering
  \includegraphics[width=0.5\textwidth]{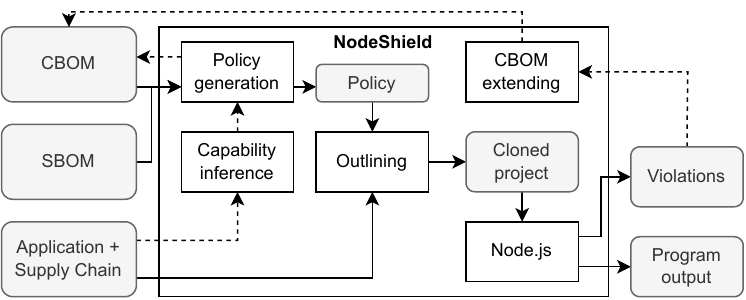}
  \caption{Architecture and Workflow of NodeShield.}
  \label{figure:workflow}
\end{figure}

\subsection{Enforcement}
\label{section:implementation:enforcement}

NodeShield implements the policy enforcement by managing the global namespace per module, intercepting any import attempts by the program, controlling dynamic code evaluation, and managing bindings access. This is achieved by utilizing only mechanisms provided by Node.js and the \texttt{vm} API, with no modifications to Node.js.

\tightpar{Global variables}
To manage the global namespace of all modules, we define specialized sets of global variables \new{(line 17-19, \Cref{listing:outlining})} and utilize the option to specify a ``context'' object when using \texttt{vm} \new{(line 25, \Cref{listing:outlining})}. By default, only standard JavaScript built-ins (e.g., \texttt{Object} or \texttt{Error}) are available. Specifying a context object allows for extending and overriding the namespace through the properties of the context object which are accessible as plain identifiers in the evaluated script. Thus, by omitting capability-bearing global variables that the current module should not have access to we can prevent it from using that capability.

Because we control the creation of all modules, we can control their global variables as specified in the CBOM. The set of global variables made available to each module starts from the set of all global variables in the host context, excluding capability-bearing global variables \new{(line 17, \Cref{listing:outlining})}. This set is extended with the allowed capability-bearing global variables of the module, according to the CBOM \new{(line 18, \Cref{listing:outlining})}.

\tightpar{Imports}
To comprehensively intercept all import attempts we need to cover all three import options: \texttt{require} for \commonjs, \texttt{import} declarations for \esmodules, and the \texttt{import} function for both. The \texttt{require} function must be provided through the global namespace \new{(line 21, \Cref{listing:outlining})}, following the above description. As such, we can intercept all calls to \texttt{require}. For \texttt{import} declarations \texttt{vm} requires an implementation of a resolution algorithm through its \texttt{link} API when instantiating a (\esmodules) module \new{(line 27, \Cref{listing:outlining})}. Hence, we intercept all uses of the \texttt{import} syntax. Finally, for the \texttt{import} function to work, \texttt{vm} also requires an implementation of a resolution algorithm through the \texttt{importModuleDynamically} API \new{(line 26, \Cref{listing:outlining})} when instantiating either a (\commonjs) script or (\esmodules) module. As a result, we also intercept all uses of the \texttt{import} function.

A fourth consideration is the built-in module named \texttt{module} which has a \texttt{createRequire} function. This function can be used to create a new \texttt{require} function which could bypass our enforcement. Because accessing this function requires importing \texttt{module}, following the previous paragraph, we can intercept all attempts to import \texttt{module}. We leverage this to return a modified \texttt{module} interface where the \texttt{createRequire} function creates \texttt{require} functions that perform the import policy check.

Given our ability to intercept all import attempts by the application, we can enforce dependency imports according to the SBOM by allowing an import if the specifier matches that of a listed dependency. Similarly, we can enforce the CBOM by allowing imports of capability-bearing built-in modules only if the corresponding capability is granted. Additionally, the import is allowed if it is for a file within the current package or an unprivileged built-in module.

\tightpar{Code evaluation}
To manage the ability of modules to dynamically evaluate code we use the \texttt{codeGeneration} option of the \texttt{vm} module. This option tells \texttt{vm} whether or not to allow dynamic text- or WASM-based code evaluation. Disabling the former disables \texttt{eval} (and related), while disabling the latter disables the functionality of the \texttt{WebAssembly} object. In both cases, use of these APIs results in an immediate error when called.

Since we control the creation of all modules, we can control the dynamic code evaluation capability for every module in accordance with the CBOM. Observe that even though \texttt{vm} can be used to create a new context where dynamic code evaluation is possible (in fact, NodeShield relies on this fact), the ability to import \texttt{vm} is guarded by the same capability as dynamic code evaluation, meaning it cannot be used to escalate capabilities.

\tightpar{Bindings}
To manage the binding accessible to each module we need to manage the behavior of \texttt{process.binding}. The \texttt{process} object is accessible both as a global variable (named \texttt{process}) and an import (as \texttt{(node:)process}). As described, we control access to both global variables and imports. Hence, we override the \texttt{binding} function on the \texttt{process} object \new{(line 15, \Cref{listing:outlining})} on a per-module basis with an implementation that restricts access to capability-bearing bindings in accordance with the CBOM.

\tightpar{Enforcement modes}
NodeShield supports 3 enforcement modes. First, in \texttt{log} mode it will log policy violations, along with the violating module, but otherwise continue as usual - thus not preventing the violation. Second, in \texttt{throw} mode violations will result in an error being thrown. This provides an option for the program to continue in spite of the limitation imposed on it. Lastly, the \texttt{exit} mode will result in an immediate program exit upon the first violation.

\subsection{Security Analysis}
\label{section:implementation:security-analysis}

We present a security analysis of NodeShield through the lens of attack scenarios that aim to bypass the enforcement and argue for its robustness against these attacks under explicit assumptions. Robustness is further evaluated empirically in \Cref{section:evaluation:robustness}.

The enforcement as described in \Cref{section:implementation:enforcement} is comprehensive in that it covers all avenues through which Node.js code can use third-party libraries or access capability-bearing APIs. However, Node.js has several features related to meta-programming and dynamic code evaluation that could allow advanced malware to bypass the enforcement.
For example, a naive implementation of the enforcement could be bypassed by monkey patching, e.g., the prototype of JavaScript built-ins such as \texttt{Object} or \texttt{Array}. Moreover, attacker can perform cross-V8-context code evaluation to achieve privilege escalation when evaluating in a more privileged context.

\tightpar{Assumptions}
\new{The security of NodeShield is predicated on the following assumptions on JavaScript, Node.js, and the \texttt{vm} module. First, dynamic code evaluation in JavaScript is limited to accessing only variables from the current scope---per the language specification~\cite{tc39}. Second, the only three ways to load modules in Node.js are the \texttt{require} function (in \commonjs), the import syntax (in \esmodules), and the \texttt{import} function (both)---per the Node.js docs~\cite{node-docs}. Third, we assume the \texttt{vm} module 1) traps all uses of both the import syntax and function, 2) prevents use of dynamic code evaluation APIs and 3) provides a fresh JavaScript context without Node.js-specific built-ins or implicit access to variables from the context in which it is instantiated---the former two are supported by the Node.js docs~\cite{node-docs} while we empirically validate the latter.}

\tightpar{Shared global namespace}
Without separation of the global namespace between the host and guest code, the guest code could manipulate the behavior of the host by changing built-ins or performing prototype pollution. Rather than trying to contain the guest code, we write the host using techniques that prevent such influence. In particular, we leverage primordials, \texttt{null} prototypes, and object freezing.

First, we use primordials \new{(line 2, \Cref{listing:outlining})}, a concept borrowed from Node.js \cite{primordials}. Primordials are a set of function references obtained before any untrusted code is run such that untrusted code cannot manipulate these references. For example, our policy enforcement for imports needs to check if the import specifier is present in the allowlist of imports. We implement it as a JavaScript array with the goal of using \texttt{l.includes(s)} to check if the specifier \texttt{s} is allowed to be imported. However, the guest code could override \texttt{Array.prototype.includes} to always return true, invalidating import-based policy enforcement. To prevent this attack, we obtain a primordial reference to the \texttt{includes} function\footnote{as \texttt{Function.prototype.call.bind(Array.prototype.includes)}. Used as \texttt{includes(allowlist, specifier)}.} and use it instead. Additionally, as a defense-in-depth strategy, we leverage \texttt{null} prototypes and object freezing. The use of \texttt{null} prototype avoids prototype pollution gadgets in our implementation, while object freezing avoids manipulating in-memory policy objects.

This prevents exploitation of the host through the global namespace, but not other (more privileged) guests. Indeed, changes to the \texttt{globalThis} object or standard object prototypes could be used by the attacker to trigger gadgets~\cite{GHunter2024} in other guests. The reasoning for this gap is twofold. First, modifications in the global namespace happen in benign use cases such as polyfills, which we do not want to break. Second, known defenses against this type of attack, e.g., object freezing, can be implemented at the application level in a way that is compatible with such benign use cases (which we cannot do in NodeShield).

\tightpar{Host-context eval}
The use of \texttt{vm} introduces the possibility of dynamic code evaluation in the host's context, potentially allowing malicious code to bypass the policy enforcement. We propose a novel technique based on lexical scoping of variables to mitigate privilege escalation in this manner. In particular we prune sensitive global variables (i.e., those from \Cref{table:capabilities} plus all \commonjs~specific global variables) from the global namespace of the host context, capturing them as local variables instead and \texttt{delete}ing them from the \texttt{globalThis} object afterwards \new{(line 3-4, \Cref{listing:outlining})}.

This relies on the observation that cross-context code evaluation does not allow the evaluator to access local variables of the evaluatee. Thus by pruning sensitive global variables the evaluatee does not gain additional capabilities by evaluation code in the host context.

\tightpar{Cross-V8-context eval}
A similar situation may occur when dynamic code is evaluated in another (more privileged) guest context. Here too, rather than trying to prevent cross-V8-context code evaluation, we use the same scoping-based technique to prevent such evaluation from enabling privilege escalation.

To achieve this we create one-time-accessible global variables which are bound to a local variable in the guest context through an inserted preamble. To create a one-time-accessible global variable we define a property with a \texttt{get}ter that deletes itself.
\begin{lstlisting}
Object.defineProperty(G, "fetch$2eb2a5", {
  configurable: true, get() {
    delete G["fetch$2eb2a5"]; return fetch } })
\end{lstlisting}
The inserted preamble binds the one-time-accessible global variables to the correct local names.
\begin{lstlisting}
{ let fetch = fetch$2eb2a5        // preamble
{ fetch("http://example.com") }}  // original code
\end{lstlisting}
From the perspective of the guest, accessing this local variable behaves just like the original global variable. But, because it is a local variable, it is not accessible in cross-V8-context code evaluation. Neither is the original global variable because its access is removed from the context's global namespace.

For \commonjs, we wrap the guest code in a block statement which in turn is wrapped in a block statement containing a (\texttt{let}) binding of the sensitive global variables to local variables, like the example (this approach prevents syntax errors due to re-used identifiers).
For \esmodules, we cannot wrap the entire guest code in any kind of block statement (because \texttt{import}s and \texttt{export}s must be at the top level) so we instead insert a top level statement that (\texttt{let}) binds sensitive global variables to local variables. To avoid syntax errors due to duplicate identifiers, \esmodules~code is parsed to determine and omit top-level names that are already defined.\footnote{In both cases, adding this preamble accounts for file ``headers'' such as shebangs and \texttt{"use strict";}.}

\subsection{Capability Inference and Presentation}
\label{section:implementation:capability-inference}

\begin{figure}[t]
  \centering
  \includegraphics[width=0.45\textwidth]{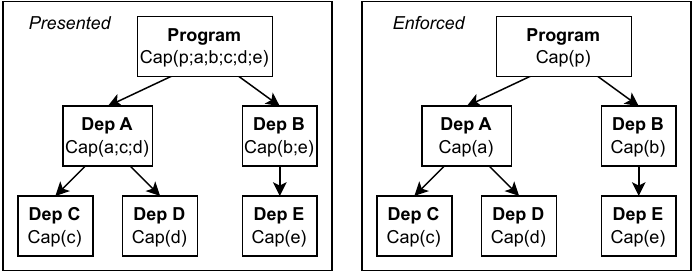}
  \caption{Presented vs Enforced Policy in NodeShield.}
  \label{figure:presented-enforced}
\end{figure}

In the absence of a CBOM a set of capabilities needs to be determined. NodeShield offers two strategies with different benefits and drawbacks. First, capabilities can be inferred dynamically by running the application using an incomplete CBOM. From the observed violations, gaps in the CBOM can be filled in. However, this risks running malware without restrictions. Second, capabilities can be inferred statically. This is imprecise given the challenges of analyzing JavaScript code, but is safer because no untrusted code is run.

\tightpar{Static inference} To statically infer the capabilities, all source code must be analyzed looking for imports of Node.js built-in modules, imports of addons, use of capability-bearing global variables, and use of capability-bearing bindings. For the prototype implementation of NodeShield we use regular expressions to match for \texttt{import}ing or \texttt{require}ing built-in modules and the presence of the word ``fetch'', ``eval'', and ``process''. This is imprecise but largely sufficient for the purposes of our evaluation. In particular, it may falsely grant capabilities if, e.g., these patterns occur in a string, or miss capabilities if, e.g., they are used dynamically only.

\tightpar{Dynamic inference} \changed{To dynamically infer capabilities, the application must be run in \texttt{log} mode, or iteratively in \texttt{exit} mode, to capture all policy violations. The policy violations can be inspected for capability-related violations, based on which the CBOM can be created or extended. Because of our use of the \texttt{codeGeneration} option, \texttt{log} mode cannot infer usage of the \emph{code} capability, requiring \texttt{exit} mode to be used instead. The prototype implementation of NodeShield does not provide an automated mechanism to create/update the CBOM in this way.}

\tightpar{Enforced vs. presented} Capabilities are enforced at a module-granular level, meaning if dependency $D_1$ requires the capability $C_1$ it receives the ability to use exactly and only that capability. When we introduce $D_2$, requiring some capability $C_2 (\neq C_1)$ as a dependency of $D_1$, a potential confused deputy problem arises: $D_1$ can use $D_2$ (in accordance with SBOM enforcement) to exercise capability $C_2$ (in violation with the CBOM enforcement). We address this by separating the \emph{enforced} (as described) and \emph{presented} view of the CBOM, illustrated in \Cref{figure:presented-enforced}.

The presented view aims to resolve the confused deputy problem (arising from the dependency hierarchy) at the moment the human reviews the CBOM while maintaining least-privilege enforcement at runtime (a motivating example for this can be found in \Cref{section:evaluation:vulnerabilities}). In particular, NodeShield users are presented a view of the dependency in which it will be granted the capabilities it needs itself as well as the capabilities of all its transitive dependencies.
\new{This leaves no deputies with extra capabilities accessible to a malicious dependency for it to confuse.} 

For the example above, the \emph{enforced} view of $D_1$ is $\{ C_1 \}$ while the presented view is $\{ C_1, C_2 \}$. This is because the code of $D_1$ itself only requires $C_1$ to function correctly---hence there is no need to grant it additional capabilities at runtime---while the overall behavior of the code comprising $D_1$ (i.e., including $D_2$) requires both $C_1$ and $C_2$ to function correctly.

\section{Evaluation}
\label{section:evaluation}

This section presents the evaluation of NodeShield~\cite{artifact}, answering the following research questions:

\begin{enumerate}[label=RQ\arabic*.]
  \item Is NodeShield effective at preventing software supply chain attacks?
  \item Is NodeShield effective at reducing the attack surface of dependencies?
  \item Is NodeShield robust against sandbox breakouts?
  \item What is the level of effort required to maintain a CBOM?
  \item What is the cost of using NodeShield compared to Node.js?
\end{enumerate}

\new{We compare NodeShield to related work~\cite{ferreira2021containing,ohm2023you} on these questions. The tool from~\cite{ferreira2021containing} is not available in a working state, hence we omit it. The Npm Dependency Guardian (ndg) tool~\cite{ohm2023you} is available and we compare with it in RQ1, RQ2, and RQ3.}

\paragraph{Experimental setup}
All experiments were conducted on a desktop system with an AMD Ryzen\texttrademark~7 3700X $\times$ 16 processor and 32~GB of RAM. For each experiment, the setup and supporting material is available in the artifact, except for the malicious code samples which are only made available upon request.
All SBOMs were generated using the \verb|npm sbom| utility.

\subsection{RQ1: Effectiveness against Malware}
\label{section:evaluation:malware}

To evaluate the effectiveness of NodeShield against software supply chain attacks we run known malicious \npm~packages and observe whether the attack can be stopped. In particular, we investigate whether the attack is detected by NodeShield through either SBOM or CBOM enforcement, and compare to ndg~\cite{ohm2023you}.

We compile a list of known malicious packages from prior work~\cite{duan2020maloss,ohm2020backstabber,geer2020good,huang2024donapi} and gray literature~\cite{cncfcatalog,openssfcasestudies}. We obtain samples of the malicious code from~\cite{duan2020maloss,ohm2020backstabber} as well as Socket (\url{https://socket.dev/}). In total, we collect the names of 2,179 known malicious packages (with one or more version) with samples available for (at least one malicious version of) 2,101 packages. Out of this, we exclude 407 unlabeled packages, 49 web platform attacks, 4 denial of service attacks, 2 malicious \texttt{npm test} commands, 2 unpublished packages, 1 that is not itself malicious, and 1 that is not designed for Node.js. This leaves 1,594 packages labeled as install-time and 41 packages labeled as runtime attacks.

First, we consider the 41 packages with a runtime attack. These can be split into two categories, malicious \emph{updates} (n=4) and malicious \emph{packages} (n=37). For this experiment we create a program that imports (and uses, if necessary) the malicious package. For malicious \emph{updates} we run the program with NodeShield using an SBOM generated for the program using the malicious package version and a statically inferred CBOM (\Cref{section:implementation:capability-inference}) for the program using the previous benign package version (extended with inferred capabilities for new transitive dependencies of the malicious version). \new{Similarly, we test ndg with an inferred policy for the benign version updated to include new dependencies from the malicious version.} The results are summarized in \Cref{table:evaluation:malicious-updates}.

For malicious \emph{packages} we run the program with NodeShield using an SBOM generated for the program with the malicious package version and using 1) a statically inferred CBOM for the program using the malicious package version, and separately 2) a CBOM that grants the malicious package no capabilities. \new{For ndg we run with the inferred policy only, assuming the results of test 2 apply to ndg too.} The results are summarized in \Cref{table:evaluation:malicious-packages}.

\begin{table}
  \rowcolors{2}{white}{gray!20}
  \begin{tabularx}{\columnwidth}{ L{0.409\columnwidth} | C{0.14\columnwidth} | C{0.14\columnwidth} | C{0.14\columnwidth} }
    \hline
    \textbf{Package}       & \textbf{SBOM} & \textbf{CBOM} & \new{\textbf{ndg}~\cite{ohm2023you}} \\
    \hline
    conventional-changelog & \emptycirc    & \fullcirc     & \new{\fullcirc}                      \\
    event-stream           & \emptycirc    & \fullcirc     & \new{\fullcirc}                      \\ 
    node-ipc               & \emptycirc    & \emptycirc    & \new{\emptycirc}                     \\
    rate-map               & \fullcirc     & \fullcirc     & \new{\fullcirc}                      \\
    \hline
  \end{tabularx}
  \caption{
    Overview of using NodeShield and ndg with malicious \textit{updates}.
    A \fullcirc~means it is prevented using that enforcement alone, \emptycirc~means it is not.
  }
  \label{table:evaluation:malicious-updates}
\end{table}

\changed{Second, we consider 5 packages with an install-time attack. We restrict ourselves to 5 out of 1,594 samples, matching coverage of related work~\cite{ohm2023you,ferreira2021containing}, because the evaluation requires manual setup to run the scripts explicitly (as opposed to implicitly through \npm~hooks). Further, we note that installation scripts are not limited to JavaScript code and we advocate instead for mechanisms along the lines of Latch~\cite{wyss2022wolf} or LavaMoat's~\cite{lavamoat} \texttt{@lavamoat/allow-scripts} tool which provide more comprehensive defenses against install-time attacks.}

All 5 packages are malicious \emph{updates}. For this experiment we create a program that runs the install time script with NodeShield using an SBOM generated for the program using the malicious package version and using 1) a statically inferred CBOM for the program using the previous benign package version, and separately 2) a statically inferred CBOM for the program using the previous benign package version where the package is stripped of all non-install-time code. \new{For ndg we run the same two experiments with inferred policies.} The results are summarized in \Cref{table:evaluation:malicious-install}.

\begin{table}
  \rowcolors{2}{gray!20}{white}
  \begin{tabularx}{\columnwidth}{ L{0.409\columnwidth} | C{0.14\columnwidth} | C{0.14\columnwidth} | C{0.14\columnwidth} }
    \hline
    \textbf{Package}           & \textbf{SBOM} & \textbf{CBOM} & \new{\textbf{ndg}~\cite{ohm2023you}} \\
    \hline
    @roku-web-core/ajax (2x)   & \emptycirc    & \halfcirc     & \new{\halfcirc}                      \\
    alicon                     & \emptycirc    & \halfcirc     & \new{\halfcirc}                      \\
    bb-builder                 & \emptycirc    & \halfcirc     & \new{\halfcirc}                      \\
    bitcionjslib (2x)          & \emptycirc    & \halfcirc     & \new{\xmark}                         \\
    bitcoisnj-lib (2x)         & \emptycirc    & \halfcirc     & \new{\xmark}                         \\
    botbait (3x)               & \emptycirc    & \halfcirc     & \new{\halfcirc}                      \\
    chalc                      & \emptycirc    & \halfcirc     & \new{\halfcirc}                      \\
    colorsss (2x)              & \emptycirc    & \halfcirc     & \new{\halfcirc}                      \\
    colors\_express            & \emptycirc    & \fullcirc     & \new{\fullcirc}                      \\
    commender                  & \emptycirc    & \halfcirc     & \new{\halfcirc}                      \\
    component-emiter           & \emptycirc    & \halfcirc     & \new{\halfcirc}                      \\
    discord-fix                & \fullcirc     & \halfcirc     & \new{\halfcirc}                      \\
    discord-lofy               & \emptycirc    & \halfcirc     & \new{\halfcirc}                      \\
    discord-selfbot-v14        & \emptycirc    & \halfcirc     & \new{\halfcirc}                      \\
    discord-vilao              & \fullcirc     & \fullcirc     & \new{\halfcirc}                      \\
    discord.js-user            & \emptycirc    & \halfcirc     & \new{\halfcirc}                      \\
    discordi.js (7x)           & \emptycirc    & \halfcirc     & \new{\halfcirc}                      \\
    discordsystem              & \emptycirc    & \halfcirc     & \new{\halfcirc}                      \\
    electron-native-notify     & \emptycirc    & \fullcirc     & \new{\fullcirc}                      \\ 
    esprime                    & \fullcirc     & \halfcirc     & \new{\xmark}                         \\
    express-cookies            & \emptycirc    & \fullcirc     & \new{\xmark}                         \\ 
    fast-requests              & \fullcirc     & \emptycirc    & \new{\emptycirc}                     \\
    flatmap-stream             & \emptycirc    & \fullcirc     & \new{\fullcirc}                      \\
    getcookies                 & \emptycirc    & \fullcirc     & \new{\xmark}                         \\
    headcache                  & \emptycirc    & \halfcirc     & \new{\halfcirc}                      \\
    http-proxy-middelware      & \emptycirc    & \halfcirc     & \new{\halfcirc}                      \\
    ikst                       & \emptycirc    & \halfcirc     & \new{\halfcirc}                      \\
    jquerry                    & \fullcirc     & \halfcirc     & \new{\halfcirc}                      \\
    leetlog (2x)               & \emptycirc    & \halfcirc     & \new{\halfcirc}                      \\
    mendiff                    & \emptycirc    & \halfcirc     & \new{\halfcirc}                      \\
    momnet (4x)                & \emptycirc    & \halfcirc     & \new{\halfcirc}                      \\
    monent                     & \emptycirc    & \halfcirc     & \new{\halfcirc}                      \\
    npmpubman                  & \emptycirc    & \halfcirc     & \new{\halfcirc}                      \\
    prerequests-xcode          & \emptycirc    & \halfcirc     & \new{\halfcirc}                      \\
    random-vo...generator (4x) & \emptycirc    & \halfcirc     & \new{\halfcirc}                      \\
    seemver                    & \emptycirc    & \halfcirc     & \new{\halfcirc}                      \\
    stautses                   & \emptycirc    & \halfcirc     & \new{\halfcirc}                      \\
    \hline
  \end{tabularx}
  \caption{
    Overview of using NodeShield and ndg with malicious \textit{packages}.
    \fullcirc~means it is prevented using that enforcement alone, \emptycirc~means it is not.
    For \textit{CBOM}, \halfcirc~means it can be prevented if the capabilities for the package are restricted w.r.t. those inferred.
    A \xmark~means there is a compatibility issue.
  }
  \label{table:evaluation:malicious-packages}
\end{table}

\tightpar{Results}
We evaluate NodeShield against 67 malware samples (46 malicious packages). The results show that our approach works best for malicious \emph{updates} and can work for malicious \emph{packages} provided developers review the capabilities before adopting a new dependency. We find that SBOM enforcement by itself is rarely sufficient to prevent known attacks, but is occasionally necessary (e.g., \texttt{fast-requests}). The results confirm the benefits of the CBOM in addition to the SBOM. In response to RQ1, we find that NodeShield could have prevented 98.51\% (66/67) of evaluated known supply chain attacks. \new{In contrast, ndg~\cite{ohm2023you} could have prevented 83.58\% (56/67) attacks. The fact that 9 attack are incompatible with ndg motivates the comprehensive and runtime-agnostic approach of NodeShield.}

The \texttt{node-ipc} attack is not prevented by NodeShield because the preceding benign version already uses the capabilities (\texttt{file-system} and \texttt{network}) used in the attack. Compared to related work~\cite{ohm2023you,ferreira2021containing}, our evaluation covers a strict superset of in-scope (i.e., excluding availability attacks) malicious packages. Furthermore, the permission system of Ferreira et al.~\cite{ferreira2021containing} only covers the \texttt{network}, \texttt{file-system}, and \texttt{command} capability\footnote{Unfortunately, the prototype of Ferreira et al. is no longer supported, as confirmed by personal communication, hence an empirical comparison is not possible.}, while NodeShield demonstrates the need for including the capabilities \texttt{system} and \texttt{crypto}. In particular, 10 samples only use one or both of these internally while relying on third-party modules for further capabilities.

\begin{table}
  \rowcolors{2}{gray!20}{white}
  \begin{tabularx}{\columnwidth}{ L{0.409\columnwidth} | C{0.14\columnwidth} | C{0.14\columnwidth} | C{0.14\columnwidth} }
    \hline
    \textbf{Package}     & \textbf{SBOM} & \textbf{CBOM} & \new{\textbf{ndg}~\cite{ohm2023you}} \\
    \hline
    eslint-config-eslint & \emptycirc    & \fullcirc     & \new{\fullcirc}                      \\
    eslint-scope         & \emptycirc    & \fullcirc     & \new{\fullcirc}                      \\
    kraken-api           & \emptycirc    & \fullcirc     & \new{\fullcirc}                      \\
    mariadb              & \emptycirc    & \halfcirc     & \new{\xmark}                         \\
    opencv.js            & \emptycirc    & \fullcirc     & \new{\xmark}                         \\
    \hline
  \end{tabularx}
  \caption{
    Overview of using NodeShield and ndg with malicious \textit{install-time} package.
    \fullcirc~means it is prevented using that enforcement alone, \emptycirc~means it is not.
    For \textit{CBOM}, \halfcirc~means the attack can be prevented if the installation script has separate capabilities.
    A \xmark~means there is a compatibility issue.
  }
  \label{table:evaluation:malicious-install}
\end{table}

\subsection{RQ2: Attack Surface Reduction}
\label{section:evaluation:vulnerabilities}

\changed{To evaluate the reduction of the attack surface we run known vulnerability exploits on NodeShield, as well as ndg~\cite{ohm2023you} for comparison, to see if the exploit is caught.} For this evaluation we use SecBench.js~\cite{bhuiyan2023secbench} (at commit \verb|bc31562|), which provides proof of concept (PoC) exploits for code injection, command injection, path traversal, prototype pollution, and ReDoS vulnerabilities.

For this experiment, only code injection vulnerabilities are within the scope of NodeShield. In particular, command injection is subject only to the \texttt{command} capability, path traversal is subject only to the \texttt{file-system} capability, and prototype pollution and ReDoS are out of scope. On the other hand, code injection enables the attacker to leverage further capabilities.

Hence, we take the PoC code injection exploits from SecBench.js and put each in a separate project that has a dependency on the vulnerable version of the respective dependency. For NodeShield, we use a generated SBOM and statically inferred CBOM (\Cref{section:implementation:capability-inference}, extended with the \verb|code| and \verb|system| capability for the vulnerable package, if not inferred). \new{Similarly, for ndg we use its inferred policy.}

We exclude some cases from our evaluation because: 16 are listed but have no PoC exploit, 12 cause prototype pollution (which is outside our threat model), 3 are mislabeled (as \verb|code-injection| instead of \verb|command-injection|), and 1 has an invalid PoC exploit.
During the experiment, the programs are run and we observe if the exploits are caught as a violation by NodeShield and ndg. The results are summarized in \Cref{table:evaluation:ace}.

\tightpar{Results}
In response to RQ2, we find that NodeShield is effective at reducing the impact of arbitrary code execution vulnerabilities in JavaScript. In particular, 87.50\% (21/24) of the exploits were detected by NodeShield. This demonstrates that the reduced attack surface as a result of capability enforcement is effective at protecting against the exploitation of code injection vulnerabilities.
\new{In contrast, 12/24 (50.00\%) of exploit were detected by ndg~\cite{ohm2023you} with 8 of the packages being incompatible.}

\newer{All attacks not prevented by NodeShield leverage the capabilities needed by the respective packages. ndg prevents two attacks NodeShield does not because these access \texttt{require}, which it disallows. However, sandbox breakouts can be used to bypass this protection (demonstrated by the \texttt{jsen} and \texttt{mongo-parse} attacks).}

Moreover, this evaluation highlights the importance of separating the presented view from the enforced view (\Cref{section:implementation:capability-inference}). For example, the exploit for the \texttt{mobile-icon-resizer} package uses the \texttt{file-system} capability. However, some of its transitive dependencies do need this capability. If the presented view was instead used for enforcement, \texttt{mobile-icon-resizer} would have received the capability and the exploit would not have been prevented.

\begin{table}
  \rowcolors{2}{gray!20}{white}
  \begin{tabularx}{\columnwidth}{ L{0.319\columnwidth} | L{0.17\columnwidth} | C{0.19\columnwidth} | C{0.15\columnwidth} }
    \hline
    \textbf{Package}     & \textbf{Capability}   & \textbf{NodeShield} & \new{\textbf{ndg}~\cite{ohm2023you}} \\ 
    \hline
    access-policy        & file-system           & \fullcirc           & \new{\fullcirc}                      \\ 
    cd-messenger         & file-system           & \fullcirc           & \new{\fullcirc}                      \\ 
    hot-formula-parser   & command               & \fullcirc           & \new{\fullcirc}                      \\ 
    jsen                 & command               & \fullcirc           & \new{\emptycirc}                     \\ 
    json-ptr             & command               & \fullcirc           & \new{\emptycirc}                     \\ 
    kmc                  & file-system           & \emptycirc          & \new{\fullcirc}                      \\ 
    m-log                & file-system           & \fullcirc           & \new{\xmark}                         \\ 
    mathjs (2x)          & command               & \fullcirc           & \new{\xmark}                         \\ 
    mixin-pro            & file-system           & \fullcirc           & \new{\fullcirc}                      \\ 
    mobile-icon-resizer  & file-system           & \fullcirc           & \new{\xmark}                         \\ 
    modjs                & file-system           & \emptycirc          & \new{\xmark}                         \\ 
    modulify             & file-system           & \emptycirc          & \new{\fullcirc}                      \\ 
    mol-proto            & file-system           & \fullcirc           & \new{\fullcirc}                      \\ 
    mongo-parse          & command               & \fullcirc           & \new{\emptycirc}                     \\ 
    mongoosemask         & file-system           & \fullcirc           & \new{\fullcirc}                      \\ 
    node-extend          & file-system           & \fullcirc           & \new{\fullcirc}                      \\ 
    node-rules           & file-system           & \fullcirc           & \new{\xmark}                         \\ 
    node-serialize       & file-system           & \fullcirc           & \new{\fullcirc}                      \\ 
    pixl-class           & file-system           & \fullcirc           & \new{\fullcirc}                      \\ 
    reduce-css-calc      & file-system           & \fullcirc           & \new{\fullcirc}                      \\ 
    serialize-to-js      & command               & \fullcirc           & \new{\emptycirc}                     \\ 
    thenify              & file-system           & \fullcirc           & \new{\xmark}                         \\ 
    underscore           & file-system           & \fullcirc           & \new{\xmark}                         \\ 
    \hline
  \end{tabularx}
  \caption{
    Overview of the protection of NodeShield and \textbf{ndg}~\cite{ohm2023you} against arbitrary code execution vulnerability exploits from SecBench.js~\cite{bhuiyan2023secbench}.
    The \emph{capability} column specifies which capability the exploit uses.
    A \fullcirc~means the PoC exploit is stopped, \emptycirc~means it is not.
    A \xmark~means there is a compatibility issue.
  }
  \label{table:evaluation:ace}
\end{table}

\subsection{RQ3: Robustness against Attack}
\label{section:evaluation:robustness}

To empirically evaluate the security of our enforcement strategy, as described in \Cref{section:implementation:security-analysis}, we collect a benchmark of language-level JavaScript sandbox breakouts from SandDriller~\cite{alhamdan2023sanddriller}. We test these on NodeShield, as well as ndg~\cite{ohm2023you} for comparison. We consider snippets from figures as well as snippets referenced in Table 5 in SandDriller~\cite{alhamdan2023sanddriller}. We create a separate program to execute each snippet and adapt them in a way that prevents the capability inference (\Cref{section:implementation:capability-inference}) from granting the program the capability it may be trying to obtain. Some snippets were omitted because they are not applicable (e.g., they target the browser environment). In total, we create a benchmark of 27 snippets.

For NodeShield, we run the programs using an SBOM generated for the program and a statically inferred CBOM. For~\cite{ohm2023you}, node-dependency-guardian (ndg), we run the program under a policy that disallows everything except access to the global variables \texttt{Buffer}, \texttt{console}, \texttt{Error}, \texttt{Object}, \texttt{process}, \texttt{require}, and \texttt{setTimeout}. The results are summarized in \Cref{table:evaluation:sandbox}.

\tightpar{Results}
In response to RQ3, we find that NodeShield can handle all sandbox breakout techniques in the benchmark except for those that only achieve prototype pollution (2/29, 6.90\%), which is outside our threat model. In contrast, 11/29 (37.93\%) sandbox breakouts work on ndg~\cite{ohm2023you}, allowing advanced malware to bypass its enforcement.

\begin{table}
  \rowcolors{2}{gray!20}{white}
  \begin{tabularx}{\columnwidth}{ L{0.473\columnwidth} | C{0.20\columnwidth} | C{0.20\columnwidth} }
    \hline
    \textbf{Index}            & \textbf{NodeShield}  & \textbf{ndg}~\cite{ohm2023you} \\
    \hline
    Figure 1 (CVE-2021-23449) & \fullcirc            & \emptycirc                     \\
    Figure 5                  & \emptycirc           & \emptycirc                     \\
    Figure 6                  & \fullcirc            & \fullcirc                      \\
    Figure 7                  & \fullcirc            & \fullcirc                      \\
    Figure 8                  & \emptycirc           & \emptycirc                     \\
    vm2 issue \#138           & \fullcirc            & \emptycirc                     \\
    vm2 issue \#175           & \fullcirc            & \fullcirc                      \\
    vm2 issue \#177           & \fullcirc            & \fullcirc                      \\
    vm2 issue \#179           & \fullcirc            & \fullcirc                      \\
    vm2 issue \#184           & \fullcirc            & \emptycirc                     \\
    vm2 issue \#185           & \fullcirc            & \fullcirc                      \\
    vm2 issue \#186           & \fullcirc            & \fullcirc                      \\
    vm2 issue \#187           & \fullcirc            & \fullcirc                      \\
    vm2 issue \#197           & \fullcirc            & \emptycirc                     \\
    vm2 issue \#199           & \fullcirc            & \emptycirc                     \\
    vm2 issue \#224           & \fullcirc            & \fullcirc                      \\
    vm2 issue \#225           & \fullcirc            & \fullcirc                      \\
    vm2 issue \#241           & \fullcirc            & \fullcirc                      \\
    vm2 issue \#268           & \fullcirc            & \fullcirc                      \\
    vm2 issue \#276           & \fullcirc            & \emptycirc                     \\
    safe-eval issue \#5       & \fullcirc            & \emptycirc                     \\
    safe-eval issue \#16      & \fullcirc            & \emptycirc                     \\
    safe-eval issue \#18      & \fullcirc            & \emptycirc                     \\
    safe-eval issue \#19      & \fullcirc            & \fullcirc                      \\
    safe-eval issue \#24 (1)  & \fullcirc            & \fullcirc                      \\
    safe-eval issue \#24 (2)  & \fullcirc            & \fullcirc                      \\
    Michał Bentkowski (1)     & \fullcirc            & \fullcirc                      \\
    Michał Bentkowski (2)     & \fullcirc            & \fullcirc                      \\
    Michał Bentkowski (3)     & \fullcirc            & \fullcirc                      \\
    \hline
  \end{tabularx}
  \caption{
    Overview of the impact of sandbox breakouts from SandDriller~\cite{alhamdan2023sanddriller} on NodeShield and \textbf{ndg}. A \fullcirc~means the breakout fails and a \emptycirc~means it succeeds.
  }
  \label{table:evaluation:sandbox}
\end{table}

\subsection{RQ4: Maintenance Effort}
\label{section:evaluation:maintenance}

\changed{To get an indication of the maintenance required for using NodeShield we evaluate the CBOM size and frequency of CBOM changes. Our focus is on the CBOM because developers need to put in no (or little, see \Cref{section:limitations}) manual effort to use NodeShield.}

\new{To evaluate size, we consider CBOMs generated across all evaluations. For the evaluation of NodeShield we have generated 143 CBOMs. The average CBOM spans 78 dependencies with 64 capabilities total, or 0.82 capability per dependency. We note the relative size may be an overestimate as RQ1 and RQ2 target packages with capabilities.}

To evaluate change frequency, we use 6 git-based server projects, picked from the evaluations of related work~\cite{de2014nodesentry,wang2023hodor}, and consider the 1,000 most recent commits.
\newer{These are typically server-side Node.js projects and are thus expected to accurately capture real-world capability change patterns.}
The experiment ignores commits for which \npm~dependency installation or SBOM generation fails and terminates early if there are fewer commits. The experiment computes the total, ``reviewable'', and ``updated'' number of capability changes. \emph{Reviewable} covers new capabilities of added or updated dependencies while \emph{updated} covers only those of updated dependencies. The total is included only for completeness as removed capabilities entail no maintenance work. The results are in \Cref{table:evaluation:maintenance}.

\tightpar{Results}
\changed{In response to RQ4, we find that projects can expect fewer capabilities than dependencies and that capability updates are infrequent.} For dependency updates, application developers need to review less than 1 capability per dependency-changing commit on average. When including new dependencies, this ranges from 0.67 and 12.81 capabilities per dependency-changing commit, highlighting the difference in maintenance work between new and existing dependencies. The variance suggests that careful selection of dependencies can reduce the review workload. In conclusion, maintaining a CBOM for an application can help protect against supply chain attacks at a low overhead. This is in line with findings of related work~\cite{ohm2023you,ferreira2021containing}.

\begin{table}
  \rowcolors{2}{white}{gray!20}
  \begin{tabularx}{\columnwidth}{ L{0.27\columnwidth} | L{0.2\columnwidth} | L{0.2\columnwidth} | L{0.16\columnwidth} }
    \hline
    \textbf{Application} & \textbf{Total}          & \textbf{Reviewable}   & \textbf{Updated}    \\
    \hline
    connect              & \changed{ 2.42    (58)} & \changed{ 1.38  (33)} & \changed{0.04  (1)} \\
    express              & \changed{ 2.67     (8)} & \changed{ 0.67   (2)} &          0.33  (1)  \\
    fastify              & \changed{ 2.22    (20)} & \changed{ 1.33  (12)} & \changed{0.00  (0)} \\
    json-server          & \changed{19.00   (285)} & \changed{ 8.53 (128)} & \changed{0.67 (10)} \\
    koa                  & \changed{25.37 (1,598)} & \changed{12.81 (807)} & \changed{0.11  (7)} \\
    st                   &           3.40    (34)  &           2.50  (25)  &          0.50  (5)  \\
    \hline
  \end{tabularx}
  \caption{
    Overview of the number of capability changes (added, removed, or updated) per project. A cell reports the average per dependency-changing-commit and sum of changes across all commit (resp.).
  }
  \label{table:evaluation:maintenance}
\end{table}

\subsection{RQ5: Performance and Compatibility}
\label{section:evaluation:performance}

To estimate the cost of adopting NodeShield for protecting applications, we evaluate the performance overhead and compatibility w.r.t. vanilla Node.js. For performance, we evaluate the performance impact in terms of the response time, memory, and throughput of long-lived server applications as well as the runtime of short-lived CLI applications. For compatibility, we report on observed incompatibilities in NodeShield across all evaluated packages and applications. \new{And lastly the false positive rate for violations}.

\tightpar{Performance}
\changed{NodeShield's performance is measured in four experiments: the response time overhead, memory overhead, and throughput of long-lived server apps, and the runtime overhead of short-lived CLI apps. The first three scenarios target our use case, while the last, based on~\cite{ferreira2021containing}, measures startup overhead. We use inferred SBOMs and CBOMs and run NodeShield in \texttt{log} mode.}

The first three experiments consider server frameworks used in the evaluations of related works~\cite{de2014nodesentry,wang2023hodor}. We instantiate server applications based on the descriptions of these papers and run these applications with Node.js and NodeShield.
First, we measure the response time overhead as the difference between Node.js and NodeShield of the average of 5,000 requests.
\new{Second, we measure the memory overhead by comparing the memory usage of the server when running on Node.js against NodeShield.}
Third, we measure the throughput (requests per second, RPS) and compare between Node.js and NodeShield. Specifically, we start by sending $n$ parallel requests and increment if all responses have been received in less than 1 second. This is repeated until handling $n$ requests takes 1 second or more, 5 times in a row. The results are in \Cref{table:evaluation:performance-server}.

For the third experiment we use short-lived CLI applications used in the evaluation of~\cite{ferreira2021containing}. We run two separate experiments. First, the applications are run as is with Node.js and NodeShield and their runtime is compared. Second, the applications are run following the methodology of~\cite{ferreira2021containing} with Node.js and NodeShield and their runtime is compared, providing a common baseline for comparison, in absence of their tool. The results are in \Cref{table:evaluation:performance-cli}.

\tightpar{Results} \changed{The experiments show that NodeShield induces low runtime overhead but noticeable memory and startup overhead. We observe a response time overhead of less than 1 ms (0.31\%-1.99\%) and we find a throughput reduction of up to 360 requests per second (0.00\%-11.84\%). Memory overhead incurred by NodeShield can vary, ranging from 42.50\% to 250.74\%. This is due to the use of \texttt{vm}, linking it to dependency count rather than workload. At 120 dependencies, \texttt{json-server} suggests a few MBs of memory may be required for Node.js apps, acceptable for modern server hardware. The startup overhead may be up to 4$\times$ (experiment A), which is because the \texttt{vm} creation and policy enforcement primarily happen during this phase (supported by experiment B).}

\begin{table*}
  \rowcolors{2}{gray!20}{white}
  \begin{tabular}{ l | r|r|r | r|r|r | r|r|r }
    \hline
                    & \multicolumn{3}{c|}{\textbf{Response Overhead} (ms)} & \multicolumn{3}{c|}{\textbf{Throughput} (RPS)}         & \multicolumn{3}{c}{\new{\textbf{Memory Overhead} (kB)}} \\
    \hline
    \textbf{Server} & Node.js         & NodeShield       & Overhead        & Node.js         & NodeShield      & Reduction          & \new{Node.js} & \new{NodeShield} & \new{Overhead}       \\
    \hline
    connect         & \changed{5.267} & \changed{5.305} & \changed{0.72\%} & \changed{8,310} & \changed{8,150} & \changed{ -1.93\%} & \new{219,344} & \new{334,512}    & \new{ 52.51\%}       \\
    express         & \changed{5.726} & \changed{5.839} & \changed{1.99\%} & \changed{3,210} & \changed{2,830} & \changed{-11.84\%} & \new{175,584} & \new{363,056}    & \new{106.77\%}       \\
    fastify         & \changed{5.364} & \changed{5.387} & \changed{0.44\%} & \changed{7,820} & \changed{7,610} & \changed{ -2.69\%} & \new{257,168} & \new{715,232}    & \new{178.11\%}       \\
    json-server     & \changed{8.208} & \changed{8.314} & \changed{1.29\%} & \changed{3,410} & \changed{3,050} & \changed{-10.56\%} & \new{175,104} & \new{614,160}    & \new{250.74\%}       \\
    koa             & \changed{5.217} & \changed{5.234} & \changed{0.31\%} & \changed{8,110} & \changed{7,830} & \changed{ -3.45\%} & \new{221,696} & \new{375,184}    & \new{ 69.23\%}       \\
    st              & \changed{5.392} & \changed{5.464} & \changed{1.34\%} & \changed{4,380} & \changed{4,380} & \changed{ -0.00\%} & \new{217,440} & \new{309,856}    & \new{ 42.50\%}       \\
    \hline
  \end{tabular}
  \caption{
    Overview of long-lived server application performance test results.
    Response overhead is the response-time overhead from the perspective of the client.
    Throughput is the number of concurrent requests the server can handle in 1 second.
    Memory overhead is the additional memory used by NodeShield.
  }
  \label{table:evaluation:performance-server}
\end{table*}

\begin{table*}
  \rowcolors{2}{gray!20}{white}
  \begin{tabular}{ l | r|r|r|r | r|r|r|r }
    \hline
                         & \multicolumn{4}{c|}{\textbf{Experiment A} (ms)}                                   & \multicolumn{4}{c}{\textbf{Experiment B} (ms)}                                    \\
    \hline
    \textbf{Application} & Node.js            & NodeShield         & \multicolumn{2}{c|}{Overhead}           & Node.js            & NodeShield         & \multicolumn{2}{c}{Overhead}            \\
    \hline
    d3-dsv               & \changed{  114.62} & \changed{  212.16} & \changed{   97.54} & \changed{ 85.10\%} & \changed{5,052.22} & \changed{5,126.35} & \changed{   74.13} & \changed{  1.47\%} \\
    docco                & \changed{  898.86} & \changed{1,201.85} & \changed{  302.89} & \changed{ 33.69\%} & \changed{5,043.12} & \changed{5,097.70} & \changed{   54.58} & \changed{  1.09\%} \\
    dot-object           & \changed{   39.90} & \changed{  120.19} & \changed{   80.29} & \changed{201.25\%} & \changed{5,047.07} & \changed{5,126.09} & \changed{   79.02} & \changed{  1.57\%} \\
    dox                  & \changed{   52.50} & \changed{  235.45} & \changed{  182.95} & \changed{348.48\%} & \changed{5,060.85} & \changed{5.239.37} & \changed{  178.53} & \changed{  3.53\%} \\
    findup               & \changed{   37.09} & \changed{   88.83} & \changed{   51.74} & \changed{139.49\%} & \changed{5,041.12} & \changed{5,093.43} & \changed{   52.31} & \changed{  1.04\%} \\
    html-minifier        & \changed{  105.51} & \changed{  446.45} & \changed{  340.94} & \changed{323.13\%} & \changed{5,100.42} & \changed{5,443.12} & \changed{  342.70} & \changed{  6.72\%} \\
    js-cfb               & \changed{   73.55} & \changed{  130.16} & \changed{   56.61} & \changed{ 76.96\%} & \changed{5,086.08} & \changed{5,136.81} & \changed{   50.73} & \changed{  1.00\%} \\
    json-refs            & \changed{  202.15} & \changed{  981.94} & \changed{  779.79} & \changed{385.75\%} & \changed{5,121.57} & \changed{5,905.51} & \changed{  783.93} & \changed{ 15.31\%} \\
    json2csv             & \changed{   51.36} & \changed{  123.92} & \changed{   72.57} & \changed{141.30\%} & \changed{5,042.24} & \changed{5,119.25} & \changed{   77.00} & \changed{  1.53\%} \\
    juice                & \changed{  381.43} & \changed{  875.99} & \changed{  494.56} & \changed{129.66\%} & \changed{5,146.21} & \changed{5,667.72} & \changed{  521.52} & \changed{ 10.13\%} \\
    metalsmith           & \changed{  144.31} & \changed{  193.65} & \changed{   49.34} & \changed{ 34.19\%} & \changed{5,114.55} & \changed{5,163.68} & \changed{   49.14} & \changed{  0.96\%} \\
    mocha                & \changed{  140.42} & \changed{  193.78} & \changed{   53.36} & \changed{ 38.00\%} & \changed{5,038.53} & \changed{5,085.31} & \changed{   46.78} & \changed{  0.92\%} \\
    mockjs               & \changed{   47.96} & \changed{  100.86} & \changed{   52.90} & \changed{110.31\%} & \changed{1,015,73} & \changed{3,101.05} & \changed{2,085.32} & \changed{205.30\%} \\
    sails                & \changed{  255.85} & \changed{1,366.53} & \changed{1,110.67} & \changed{434.11\%} & \changed{5,234.75} & \changed{6,302.20} & \changed{1,067.45} & \changed{ 20.39\%} \\
    svgicons2svgfont     & \changed{  158.56} & \changed{  280.19} & \changed{  121.63} & \changed{ 76.71\%} & \changed{5,055.82} & \changed{5,161.54} & \changed{  105.71} & \changed{  2.09\%} \\
    traceur              & \xmark             & \xmark             & \xmark             & \xmark             & \xmark             & \xmark             & \xmark             & \xmark             \\
    uglify-js            & \changed{3,406.22} & \changed{4,500.16} & \changed{1,093.94} & \changed{ 32.12\%} & \changed{5,153.41} & \changed{5,222.16} & \changed{   68.75} & \changed{  1.33\%} \\
    xss                  & \changed{   55.83} & \changed{  127.71} & \changed{   71.88} & \changed{128.74\%} & \changed{5,043.40} & \changed{5,113.82} & \changed{   70.42} & \changed{  1.40\%} \\
    yaml-front-matter    & \changed{   59.56} & \changed{  167.53} & \changed{  107.97} & \changed{181.29\%} & \changed{5,038.85} & \changed{5,091.74} & \changed{   52.89} & \changed{  1.05\%} \\
    \hline
  \end{tabular}
  \caption{
    Overview of short-lived CLI application performance test results.
    In experiment A, the application is run as is.
    In experiment B, the application is run following the approach of~\cite{ferreira2021containing}.
    For both experiments the average of 10 runs is reported.
    \texttt{metalsmith} and \texttt{svgicons2svgfont} were modified to use \texttt{Array.isArray} (instead of \texttt{instanceof Array}) for compatibility with NodeShield.
    A \xmark~means there is a compatibility issue.
  }
  \label{table:evaluation:performance-cli}
\end{table*}

\tightpar{Compatibility}
%
\changed{In total, the evaluation spans 86 real-world packages and applications, covering a total of 3,443 transitive packages. Out of 86, 3 were not compatible, giving a 96.51\% compatibility rate. 
In contrast, ndg~\cite{ohm2023you} exhibits compatibility issues with 15 out 61 real-world packages and applications tested, giving a 75.41\% compatibility rate.
For NodeShield, one incompatibility is due to the use of an undocumented API (\texttt{module.\_compile}) and two are due to the use of \texttt{instanceof Array}.}

\tightpar{Results}
We find that NodeShield is broadly compatible with software written for Node.js. While incompatible coding patterns are used in practice, they are infrequent and generally easy to overcome. We provide a more in-depth discussion about incompatibility in \Cref{section:limitations}.

\tightpar{False positive rate}
\newer{We evaluate the false positive rate of violations---i.e., how often developers can expect violations that are actually benign---by running the (assumed) benign applications of the performance evaluation. As ground truth we use the set of accesses requested at runtime. A false positive is any violation that occurs (e.g., there is a capability missing from the CBOM). Dually, a true negative is the lack of a violation when there should not be (e.g., a capability from the CBOM is being used).}

\newer{For this evaluation we use generated SBOMs and statically inferred CBOMs. To get the false positives count we run the apps---for servers sending one request---and count violations reported. Repeated violations by the same dependency are ignored. For true negatives we use an empty CBOM, count the violations reported by NodeShield, and subtract the number of false positives. Thus, true negatives (CBOM violation) are an underestimate w.r.t. to false positives (all violations), leading to a lower bound on the false positive rate.}

\newer{The results of this experiment depend on the code covered when running the applications. To give a sense of the completeness of these experiments, we measure the code coverage of all applications in the evaluation. We find an average line coverage of 15.29\% (min. 2.95\%, max. 38.93\%).}

\tightpar{Results}
\newer{Across the 24 benign applications (covering 994 transitive dependencies) we observe 387 non-violations (true negatives) and 18 violations (false positives)---7 SBOM and 11 cross-package import violations. Thus, the false positive rate ($FP \div (FP+TN)$) of NodeShield in our evaluation is 4.65\%.}

\subsection{Limitations}
\label{section:limitations}

\tightpar{Language support}
\changed{NodeShield support most of the JavaScript language and Node.js environment, empirically supported by the evaluation in \Cref{section:evaluation:performance}. However, there are three language aspects with partial support: 1) dynamic type checking, 2) overriding certain global variables, and 3) undocumented Node.js APIs.}

\changed{First, the usage of \texttt{vm} creates separate V8 contexts, each with unique built-in constructors. As a result the \texttt{instanceof} operator does not work as expected for types instantiable through syntax (e.g., arrays). This limitation can be overcome by using, e.g., \texttt{Array.isArray} instead, which is recommended practice~\cite{arrayisarray}.}

\changed{Second, our approach prevents overriding privileged global variables (from \Cref{table:capabilities}) in packages with the corresponding capability (e.g., \texttt{fetch=42}). This is due to the local binding of those variables in those packages (per \Cref{section:implementation:security-analysis}). We believe this is uncommon yet can be overcome by overriding as \texttt{globalThis.fetch=42}.}

\changed{Third, Node.js provides some APIs that are not documented. If such APIs are not implemented in NodeShield, code using these APIs breaks. We only know about \texttt{module.\_compile}. Covering such APIs is challenging as their intended functionality is unknown.}

\tightpar{Incompleteness}
Despite our systematic efforts to map all Node.js APIs to capabilities (see \Cref{table:capabilities}), we cannot prove that our mapping is complete. We cover all documented APIs and some undocumented APIs. We argue this limitation is not fundamental but rather incidental. In particular, our enforcement approach outlined in \Cref{section:implementation:enforcement} supports enforcing policies on such APIs, yet might be insufficiently configured.

\tightpar{Capabilities}
While our evaluation shows that the enhancement of SBOM with CBOM is a simple and effective abstraction, there may be challenges. First, capability overlap, most notably with \texttt{addon} and \texttt{command}, may not be obvious and potentially lead to unexpected use of resources. Second, capability transfer between components may not always be desirable. Third, as seen in \Cref{section:evaluation:malware}, when a privileged malicious packages (e.g., \texttt{node-ipc}) is compromised the attack can freely use its capabilities. We argue that many packages will not be privileged, thus NodeShield significantly reduce the attack surface. \newer{Lastly, the need for manual review of capabilities can be a limiting factor for the security gained from using NodeShield.}

\tightpar{SBOMs}
There are known issues with the accuracy of SBOMs, especially in the form of missing components. For NodeShield, this results in legitimate imports not being allowed, thus causing usability rather than security problems. Additionally, not all SBOM generators capture the dependency hierarchy, eliminating much of the benefit of SBOM enforcement from NodeShield.

\section{Malware Spotlight: Copay Wallet}
\label{section:case-study}

To further illustrate the practical benefits of NodeShield we apply it in a case study of the attack on Copay through \texttt{event-stream}~\cite{eventstreamincident}. Copay is a cryptocurrency wallet application build using Node.js and Electron. In 2018, it was targeted by a supply chain attack. A dependency of Copay, \texttt{event-stream}, was compromised to include the malicious dependency \texttt{flatmap-stream}, which contained a payload that was designed to trigger only when used by Copay.

Copay used \texttt{event-stream} as part of its development processes. It was present transitively through the use of \texttt{npm-run-all}---a utility for running multiple \npm~scripts in a convenient and concise way. The dependency hierarchy looked like:
\begin{equation*}
  \texttt{copay > npm-run-all > ps-tree > event-stream}
\end{equation*}
In the attack, \texttt{event-stream} imports \texttt{flatmap-stream} which in turn alters the files of a runtime dependency of Copay. This alteration causes Copay to leak account data and private keys of users with sufficient balance at runtime. If \texttt{npm-run-all} ran on NodeShield, the attack would be detected during the build of Copay and no backdoored versions would have been released. As such, we describe how this attack would have played out if this was the case.

For this purpose we use \texttt{npm-run-all} v4.1.2 (commit \texttt{ec4d56c}) which, due to version ranges in \texttt{ps-tree}, can depend on both a benign and malicious version of \texttt{event-stream}. We generate an SBOM and statically infer a CBOM for both cases. When using the benign version of \texttt{event-stream} (v3.3.4) the SBOM contains the above dependency relation and a CBOM with:
\begin{lstlisting}
{ "npm-run-all@4.1.2": ["system", "file-system", "command"],
  "ps-tree@1.2.0": ["system", "command"],
  "event-stream@3.3.4": ["system", "file-system"] }
\end{lstlisting}
The malicious version of \texttt{event-stream} (v3.3.6) introduces a new dependency on \texttt{flatmap-stream} through the relation:
\begin{equation*}
  \texttt{npm-run-all > ps-tree > event-stream > flatmap-stream}
\end{equation*}
and, consequently, the CBOM will also have an entry for it (but is otherwise unchanged):
\begin{lstlisting}
{ "flatmap-stream@0.1.1": ["system"] }
\end{lstlisting}

Hence, the difference between the two version comes down to the addition of the \texttt{flatmap-stream} package---as a dependency of \texttt{event-stream}---requiring access to the \texttt{system} capability. We run \texttt{npm-run-all} as a NodeShield-based project, once with v3.3.4 of \texttt{event-stream} and once with v3.3.6 of \texttt{event-stream}, showing full compatibility with the original project and preventing the attack (resp.). The attack is prevented because \texttt{flatmap-stream} uses extra capabilities (\texttt{crypto}, \texttt{file-system}, and \texttt{network}) at runtime. Notably, none of these are present in the CBOM from static capability inference because \texttt{flatmap-stream} is obfuscated. If the code was not obfuscated, these extra capabilities would appear in the CBOM, signaling the maintainers of \texttt{npm-run-all} about the change in the capabilities used by its dependencies. Considering the presented view of its direct dependencies, the update of \texttt{event-stream} adds the \texttt{network} and \texttt{crypto} capabilities to \texttt{ps-tree}. This is unexpected for a dependency that (by its own description) ``[gets] all children of a pid'', thus providing a strong indicator to review the update.

In summary, the case study shows that NodeShield would have prevented the \texttt{event-stream} incident if it was used by \texttt{npm-run-all} with no manual effort from the maintainers. Moreover, the hypothetical alternative attack (i.e., not obfuscated) would likely have been caught due to the introduction of two suspicious capabilities for an existing dependency.

\section{Related Work}
\label{section:related-work}

We discuss closely-related works and place our contributions in the broader area of web application security. NodeShield contributes with a practical open-source
system with direct focus on the supply chain of Node.js applications, while ensuring Node.js compatibility, automation, minimal overhead, and policy conciseness. To the best of our knowledge, there is no system that meets these goals.

\tightpar{Permission systems}
We are not the first to propose a system that enforces permissions on a dependency-granular level to protect against supply chain attacks. Prior work by Ferreira et al.~\cite{ferreira2021containing} and concurrent (unpublished) work by Ohm et al.~\cite{ohm2023you} propose similar systems. Our shared goal is to protect against supply chain-based malware through a system that is broadly compatible with Node.js.

Both Ferreira et al. and Ohm et al. opt to modify the Node.js runtime to trap on imports and the \texttt{globalThis} object. The former also uses source code rewriting to add dynamic property access checks. This hinders adoption because maintaining a security-enhanced fork is expensive, and while Node.js (and Deno~\cite{doglio2020introducing}) has introduced a permission system of its own~\cite{node-docs}, there is still a gap between academia and practice (e.g., these permission systems are not dependency aware). In contrast, NodeShield is built without modifying Node.js thus simplifying adoption and maintenance. Neither related work supports \esmodules, significantly hindering adoption for modern JavaScript applications. Lastly, neither consider an attacker that attempts to bypass the enforcement, which we address through lexical scoping of sensitive global variables (see \Cref{section:implementation:security-analysis}), thus preventing such attacks as shown in \Cref{section:evaluation:robustness}.

The permission system of Ferreira et al.~\cite{ferreira2021containing} covers a subset of our capabilities, missing the \texttt{system} and \texttt{crypto} capability we find used in practice as well as attacks utilizing undeclared dependencies such as the \texttt{fast-requests} package in \Cref{section:evaluation:malware}. Ohm et al.~\cite{ohm2023you} offer a more granular permission system that gives control over all imports and global variables. The resulting policy is overly verbose according to our evaluation.

From industry, LavaMoat~\cite{lavamoat} has emerged as a prominent tool to protect against JavaScript supply chain attacks. It leverage Secure ECMAScript (SES) compartments. For Node.js, it can be used as a application framework to protect against attacks from or on third-party packages. It provides stronger security guarantees (e.g., protecting against prototype pollution) at the cost of more restrictions, requiring code to be written using a subset of JavaScript.

\tightpar{Sandboxes}
More broadly, various works have looked at sandboxing for Node.js. Trading of compatibility and usability---in terms of API and policy conciseness---for security. De Groef et al.~\cite{de2014nodesentry} present NodeSentry, which uses membranes~\cite{membranes} for policy enforcement through a modified \texttt{require} implementation. Vasilakis et al.~\cite{vasilakis2018breakapp} introduce BreakApp, separating components using three isolation tiers (language, process, container) aiming to reduce vulnerability impact. They iterate on this with a language-level Read-Write-Execute-based permission model~\cite{vasilakis2021preventing} offering fine-grained control over all object properties through rewriting-based context-rebinding. Ahmadpanah et al.~\cite{ahmadpanah2021sandtrap} present a sandbox library designed for running untrusted code in Trigger-Action platforms.

The above focus primarily on isolation within the Node.js process. This leaves native extensions and subprocesses vulnerable for exploitation. Addressing these gaps, Christou et al.~\cite{christou2023binwrap} present BinWrap as a system to sandbox native code extensions along with the JavaScript itself. Similarly, Abbadini et al.~\cite{abbadini2023natisand} present NatiSand and Cage4Deno to sandbox native code extensions and subprocesses (resp.) for JavaScript runtimes using \landlock~\cite{landlock-lsm}, \ebpf~\cite{ebpf}, and \seccomp~\cite{seccomp}. Wang et al.~\cite{wang2023hodor} present HODOR, a unified system for enforcing least privilege of system call usage in Node.js applications, including native extensions, using \seccomp.

Besides runtime, the \npm~ecosystem can be subject to install-time attacks. As shown in \Cref{section:evaluation:malware}, NodeShield can be used if the script is written in JavaScript, yet installation scripts can be arbitrary scripts or programs. Wyss et al.~\cite{wyss2022wolf} propose Latch to enforce a policy on any install script by leveraging \apparmor~\cite{apparmor}. LavaMoat~\cite{lavamoat} offers tooling to manage installation scripts.

\tightpar{Web}
Besides server-side JavaScript, there has also been work on isolating client-side JavaScript. Early attempts, such as \textsc{Caja}, \textsc{ADsafe}, and \textsc{FBjs}, often relied on filtering or rewriting~\cite{maffeis2009language}. Terrace et al.~\cite{terrace2012javascript} present js.js, a JavaScript based interpreter to interpret untrusted JavaScript safely from JavaScript to achieve isolation. Agten et al.~\cite{agten2012jsand} present JSand, an SES-based sandbox that uses membranes for cross-component object sharing. Stefan et al.~\cite{stefan2014protecting} suggest COWL as an extension of browser security policies with label-based mandatory access control per script. Mickens~\cite{mickens2014pivot} present the Pivot framework for building web applications using \texttt{iframe}s as isolation containers, leveraging post messages as RPC between components.

\tightpar{Malware detection}
Many recent works analyze or detect malicious intent in the supply chain to prevent its spread. Ohm et al.~\cite{ohm2020backstabber} analyze known malicious packages for patterns and attack vectors. Ladisa et al.~\cite{ladisa2023sok} expand with a more extensive literature review and construct an attack tree covering 107 unique vectors. Similarly,~\cite{ladisa2023hitchhiker} evaluate the features of package managers to uncover attack vectors for arbitrary code execution on developer machines.

Fass et al.~\cite{fass2019jstap} propose using multiple static analyses combined with random forest classification to detect malicious JavaScript samples. Duan et al~\cite{duan2020maloss} propose using metadata, static, and dynamic analysis to detect malicious packages on \pypi, \npm, and \rubygems. Ntousakis et al.~\cite{ntousakis2021detecting} extend the work of Vasilakis et al.~\cite{vasilakis2021preventing} to detect malicious packages at runtime. Sejfia and Schäfter~\cite{sejfia2022practical} use feature-driven machine learning to detect malicious \npm~packages. Li et al.~\cite{li2023malwukong} leverage inter-procedural source-to-sink analysis to reduce false positives in detecting malicious \npm~and \pypi~packages. Liang et al.~\cite{liang2023needle} focus on detecting malicious install scripts in \pypi~packages, identifying behavioral outliers. Huang et al.~\cite{huang2024donapi} proposes the use of behavior sequences observed in known malicious packages to detect new malicious packages. Sofaer et al.~\cite{sofaer2024rogueone} focus on detecting malicious updates of packages based on changes in the use of external APIs, aligning in principle with our CBOM proposal.

\section{Conclusion}
\label{section:conclusion}

We presented a runtime protection mechanism, NodeShield, against supply chain attacks on Node.js that is able to defend against 98.51\% of tested real-world supply chain attacks and 87.50\% of tested vulnerability exploits. NodeShield requires little effort from developers and incurs low overhead on long-lived applications. Driven by a novel application of lexical scoping, NodeShield can protect against more sandbox breakouts than related works.
We applied NodeShield to a case study of the 2018 attack on the Copay application, showing its potential in practice. In future work NodeShield and CBOM can be applied in different settings with more granular policies, including other JavaScript runtimes, browsers, or languages.

\begin{acks}
We thank Daniel Hedin and the anonymous reviewers for their feedback. This work was partially supported by the Swedish Foundation for Strategic Research (SSF), the Swedish Research Council (VR),  and Wallenberg AI, Autonomous Systems and Software Program (WASP) funded by the Knut and Alice Wallenberg Foundation. 
\end{acks}

%

\bibliographystyle{plain}
\bibliography{references}

\begin{thebibliography}{10}

\bibitem{apparmor}
{AppArmor}.
\newblock \url{https://gitlab.com/apparmor/apparmor}.
\newblock Accessed: \texttt{8de2ff3}.

\bibitem{cackle}
{Cackle}.
\newblock \url{https://github.com/cackle-rs/cackle}.
\newblock Accessed: \text{6857d88}.

\bibitem{capslock}
{Capslock}.
\newblock \url{https://github.com/google/capslock}.
\newblock Accessed: \text{93953b6}.

\bibitem{ebpf}
{eBPF}.
\newblock \url{https://ebpf.io/}.
\newblock Accessed: 2024-09-13.

\bibitem{landlock-lsm}
{Landlock}.
\newblock \url{https://docs.kernel.org/security/landlock.html}.
\newblock Accessed: 2024-09-13.

\bibitem{lavamoat}
{LavaMoat}.
\newblock \url{https://github.com/LavaMoat/LavaMoat}.
\newblock Accessed: \texttt{a859f9f}.

\bibitem{seccomp}
{seccomp}.
\newblock \url{https://github.com/seccomp}.
\newblock Accessed: 2024-09-13.

\bibitem{socketalerts}
{Socket.dev} alerts.
\newblock \url{https://socket.dev/alerts}.
\newblock Accessed: 2024-09-13.

\bibitem{ossfose}
Threats, risks, and mitigations in the open source ecosystem.
\newblock \url{https://github.com/ossf/wg-metrics-and-metadata}.
\newblock Accessed: \texttt{45ff44b}.

\bibitem{abbadini2023natisand}
Marco Abbadini, Dario Facchinetti, Gianluca Oldani, Matthew Rossi, and Stefano
  Paraboschi.
\newblock {NatiSand}: Native code sandboxing for {JavaScript} runtimes.
\newblock In {\em Proceedings of the 26th International Symposium on Research
  in Attacks, Intrusions and Defenses}, pages 639--653, 2023.

\bibitem{agten2012jsand}
Pieter Agten, Steven Van~Acker, Yoran Brondsema, Phu~H Phung, Lieven Desmet,
  and Frank Piessens.
\newblock {JSand}: complete client-side sandboxing of third-party {JavaScript}
  without browser modifications.
\newblock In {\em Proceedings of the 28th Annual Computer Security Applications
  Conference}, pages 1--10, 2012.

\bibitem{ahmadpanah2021sandtrap}
Mohammad~M Ahmadpanah, Daniel Hedin, Musard Balliu, Lars~Eric Olsson, and
  Andrei Sabelfeld.
\newblock {SandTrap}: Securing {JavaScript}-driven {Trigger-Action Platforms}.
\newblock In {\em 30th USENIX Security Symposium (USENIX Security 21)}, pages
  2899--2916, 2021.

\bibitem{augur}
Mark~W. Aldrich, Alexi Turcotte, Matthew Blanco, and Frank Tip.
\newblock Augur: Dynamic taint analysis for asynchronous javascript.
\newblock In {\em Proceedings of the 37th IEEE/ACM International Conference on
  Automated Software Engineering}, ASE'22, 2023.

\bibitem{alhamdan2023sanddriller}
Abdullah Alhamdan and Cristian-Alexandru Staicu.
\newblock {SandDriller}: A fully-automated approach for testing language-based
  {JavaScript} sandboxes.
\newblock In {\em 32nd USENIX Security Symposium (USENIX Security 23)}, pages
  3457--3474, 2023.

\bibitem{bhuiyan2023secbench}
Masudul Hasan~Masud Bhuiyan, Adithya~Srinivas Parthasarathy, Nikos Vasilakis,
  Michael Pradel, and Cristian-Alexandru Staicu.
\newblock {SecBench.js}: An executable security benchmark suite for server-side
  {JavaScript}.
\newblock In {\em 2023 IEEE/ACM 45th International Conference on Software
  Engineering (ICSE)}, pages 1059--1070. IEEE, 2023.

\bibitem{cassel2023nodemedic}
Darion Cassel, Wai~Tuck Wong, and Limin Jia.
\newblock Nodemedic: End-to-end analysis of node. js vulnerabilities with
  provenance graphs.
\newblock In {\em 2023 IEEE 8th European Symposium on Security and Privacy
  (EuroS\&P)}, pages 1101--1127. IEEE, 2023.

\bibitem{christou2023binwrap}
George Christou, Grigoris Ntousakis, Eric Lahtinen, Sotiris Ioannidis,
  Vasileios~P Kemerlis, and Nikos Vasilakis.
\newblock {BinWrap}: Hybrid protection against native {Node.js} add-ons.
\newblock In {\em Proceedings of the 2023 ACM Asia Conference on Computer and
  Communications Security}, pages 429--442, 2023.

\bibitem{tc39}
ECMAScript Contributors.
\newblock {ECMAScript}® 2026 specification.
\newblock \url{https://tc39.es/ecma262/}.
\newblock Accessed: 2025-04-03.

\bibitem{arrayisarray}
MDN Contributors.
\newblock Array.{isArray}.
\newblock
  \url{https://developer.mozilla.org/en-US/docs/Web/JavaScript/Reference/Global_Objects/Array/isArray}.
\newblock Accessed: 2025-03-27.

\bibitem{node-docs}
Node.js Contributors.
\newblock Node.js v20.x documentation.
\newblock \url{https://nodejs.org/docs/latest-v20.x/api/index.html}.
\newblock Accessed: 2025-07-14.

\bibitem{primordials}
Node.js Contributors.
\newblock Usage of primordials in core.
\newblock
  \url{https://github.com/nodejs/node/blob/main/doc/contributing/primordials.md}.
\newblock Accessed: \texttt{038d829}.

\bibitem{artifact}
Eric Cornelissen and Musard Balliu.
\newblock {NodeShield}: Runtime enforcement of security-enhanced {SBOMs} for
  {Node.js} - artifact.
\newblock \url{https://doi.org/10.5281/zenodo.16873448}.

\bibitem{GHunter2024}
Eric Cornelissen, Mikhail Shcherbakov, and Musard Balliu.
\newblock Ghunter: Universal prototype pollution gadgets in javascript
  runtimes.
\newblock In {\em 33rd USENIX Security Symposium (USENIX Security 24)}, pages
  3693--3710, 2024.

\bibitem{membranes}
Tom~Van Cutsem.
\newblock {Isolating application sub-components with membranes}.
\newblock \url{https://tvcutsem.github.io/membranes}, 2018.
\newblock Accessed: 2025-03-17.

\bibitem{de2014nodesentry}
Willem De~Groef, Fabio Massacci, and Frank Piessens.
\newblock {NodeSentry}: Least-privilege library integration for server-side
  {JavaScript}.
\newblock In {\em Proceedings of the 30th Annual Computer Security Applications
  Conference}, pages 446--455, 2014.

\bibitem{doglio2020introducing}
Fernando Doglio.
\newblock Introducing {Deno}.
\newblock 2020.

\bibitem{duan2020maloss}
Ruian Duan, Omar Alrawi, Ranjita~Pai Kasturi, Ryan Elder, Brendan
  Saltaformaggio, and Wenke Lee.
\newblock Towards measuring supply chain attacks on package managers for
  interpreted languages.
\newblock 2021.

\bibitem{eslint}
ESLint.
\newblock Postmortem for malicious packages published on july 12th, 2018.
\newblock
  \url{https://eslint.org/blog/2018/07/postmortem-for-malicious-package-publishes/}.
\newblock Accessed: 2025-04-13.

\bibitem{fass2019jstap}
Aurore Fass, Michael Backes, and Ben Stock.
\newblock Jstap: A static pre-filter for malicious javascript detection.
\newblock In {\em Proceedings of the 35th Annual Computer Security Applications
  Conference}, pages 257--269, 2019.

\bibitem{ferreira2021containing}
Gabriel Ferreira, Limin Jia, Joshua Sunshine, and Christian K{\"a}stner.
\newblock Containing malicious package updates in npm with a lightweight
  permission system.
\newblock In {\em 2021 IEEE/ACM 43rd International Conference on Software
  Engineering (ICSE)}, pages 1334--1346. IEEE, 2021.

\bibitem{purescript}
Harry Garrood.
\newblock Malicious code in the {PureScript} installer.
\newblock
  \url{https://harry.garrood.me/blog/malicious-code-in-purescript-npm-installer/}.
\newblock Accessed: 2025-04-02.

\bibitem{geer2020good}
Dan Geer, Bentz Tozer, and John~Speed Meyers.
\newblock For good measure: Counting broken links: A quant’s view of software
  supply chain security.
\newblock {\em USENIX; Login}, 45(4), 2020.

\bibitem{huang2024donapi}
Cheng Huang, Nannan Wang, Ziyan Wang, Siqi Sun, Lingzi Li, Junren Chen,
  Qianchong Zhao, Jiaxuan Han, Zhen Yang, and Lei Shi.
\newblock {DONAPI}: Malicious npm packages detector using behavior sequence
  knowledge mapping.
\newblock In {\em 33rd USENIX Security Symposium (USENIX Security 24)}, pages
  3765--3782, 2024.

\bibitem{ladisa2023sok}
Piergiorgio Ladisa, Henrik Plate, Matias Martinez, and Olivier Barais.
\newblock {Sok}: Taxonomy of attacks on open-source software supply chains.
\newblock In {\em 2023 IEEE Symposium on Security and Privacy (SP)}, pages
  1509--1526. IEEE, 2023.

\bibitem{ladisa2023hitchhiker}
Piergiorgio Ladisa, Merve Sahin, Serena~Elisa Ponta, Marco Rosa, Matias
  Martinez, and Olivier Barais.
\newblock The hitchhiker's guide to malicious third-party dependencies.
\newblock In {\em Proceedings of the 2023 Workshop on Software Supply Chain
  Offensive Research and Ecosystem Defenses}, pages 65--74, 2023.

\bibitem{li2023malwukong}
Ningke Li, Shenao Wang, Mingxi Feng, Kailong Wang, Meizhen Wang, and Haoyu
  Wang.
\newblock Malwukong: Towards fast, accurate, and multilingual detection of
  malicious code poisoning in oss supply chains.
\newblock In {\em 2023 38th IEEE/ACM International Conference on Automated
  Software Engineering (ASE)}, pages 1993--2005. IEEE, 2023.

\bibitem{liang2023needle}
Wentao Liang, Xiang Ling, Jingzheng Wu, Tianyue Luo, and Yanjun Wu.
\newblock A needle is an outlier in a haystack: Hunting malicious pypi packages
  with code clustering.
\newblock In {\em 2023 38th IEEE/ACM International Conference on Automated
  Software Engineering (ASE)}, pages 307--318. IEEE, 2023.

\bibitem{maffeis2009language}
Sergio Maffeis and Ankur Taly.
\newblock Language-based isolation of untrusted javascript.
\newblock In {\em 2009 22nd IEEE Computer Security Foundations Symposium},
  pages 77--91. IEEE, 2009.

\bibitem{mickens2014pivot}
James Mickens.
\newblock {Pivot}: Fast, synchronous mashup isolation using generator chains.
\newblock In {\em 2014 IEEE Symposium on Security and Privacy}, pages 261--275.
  IEEE, 2014.

\bibitem{eventstreamincident}
npm.
\newblock Details about the event-stream incident.
\newblock
  \url{https://blog.npmjs.org/post/180565383195/details-about-the-event-stream-incident}.
\newblock Accessed: 2025-03-27.

\bibitem{getcookies}
npm.
\newblock Reported malicious module: getcookies.
\newblock
  \url{https://blog.npmjs.org/post/173526807575/reported-malicious-module-getcookies}.
\newblock Accessed: 2025-04-13.

\bibitem{ntousakis2021detecting}
Grigoris Ntousakis, Sotiris Ioannidis, and Nikos Vasilakis.
\newblock Detecting third-party library problems with combined program
  analysis.
\newblock In {\em Proceedings of the 2021 ACM SIGSAC Conference on Computer and
  Communications Security}, pages 2429--2431, 2021.

\bibitem{ohm2020backstabber}
Marc Ohm, Henrik Plate, Arnold Sykosch, and Michael Meier.
\newblock Backstabber’s knife collection: A review of open source software
  supply chain attacks.
\newblock In {\em Detection of Intrusions and Malware, and Vulnerability
  Assessment: 17th International Conference, DIMVA 2020, Lisbon, Portugal, June
  24--26, 2020, Proceedings 17}, pages 23--43. Springer, 2020.

\bibitem{ohm2023you}
Marc Ohm, Timo Pohl, and Felix Boes.
\newblock You can run but you can't hide: Runtime protection against malicious
  package updates for {Node.js}.
\newblock {\em arXiv preprint arXiv:2305.19760}, 2023.

\bibitem{Okhravi25}
Hamed Okhravi, Nathan Burow, and Fred~B. Schneider.
\newblock Software bill of materials as a proactive defense.
\newblock {\em IEEE Security \& Privacy}, 23(2):101--106, 2025.

\bibitem{SchoepeBPS16}
Daniel Schoepe, Musard Balliu, Benjamin~C. Pierce, and Andrei Sabelfeld.
\newblock Explicit secrecy: {A} policy for taint tracking.
\newblock In {\em {IEEE} European Symposium on Security and Privacy, EuroS{\&}P
  2016, Saarbr{\"{u}}cken, Germany, March 21-24, 2016}, pages 15--30, 2016.

\bibitem{sejfia2022practical}
Adriana Sejfia and Max Sch{\"a}fer.
\newblock Practical automated detection of malicious npm packages.
\newblock In {\em Proceedings of the 44th International Conference on Software
  Engineering}, pages 1681--1692, 2022.

\bibitem{jalangi}
Koushik Sen, Swaroop Kalasapur, Tasneem Brutch, and Simon Gibbs.
\newblock Jalangi: A selective record-replay and dynamic analysis framework for
  javascript.
\newblock In {\em Proceedings of the 37th IEEE/ACM International Conference on
  Automated Software Engineering}, ASE '22, 2013.

\bibitem{ShcherbakovBS23}
Mikhail Shcherbakov, Musard Balliu, and Cristian{-}Alexandru Staicu.
\newblock Silent spring: Prototype pollution leads to remote code execution in
  node.js.
\newblock In {\em 32nd {USENIX} Security Symposium, {USENIX} Security 2023,
  Anaheim, CA, USA, August 9-11, 2023}. {USENIX} Association, 2023.

\bibitem{ShcherbakovMB24}
Mikhail Shcherbakov, Paul Moosbrugger, and Musard Balliu.
\newblock Unveiling the invisible: Detection and evaluation of prototype
  pollution gadgets with dynamic taint analysis.
\newblock In {\em Proceedings of the {ACM} on Web Conference 2024, {WWW} 2024,
  Singapore, May 13-17, 2024}, pages 1800--1811, 2024.

\bibitem{snykelectron}
Snyk.
\newblock electron-native-notify malicious package.
\newblock
  \url{https://security.snyk.io/vuln/SNYK-JS-ELECTRONNATIVENOTIFY-174928}.
\newblock Accessed: 2025-04-13.

\bibitem{snykratemap}
Snyk.
\newblock rate-map malicious package.
\newblock \url{https://security.snyk.io/vuln/SNYK-JS-RATEMAP-451649}.
\newblock Accessed: 2025-04-02.

\bibitem{sofaer2024rogueone}
Raphael~J Sofaer, Yaniv David, Mingqing Kang, Jianjia Yu, Yinzhi Cao, Junfeng
  Yang, and Jason Nieh.
\newblock Rogueone: Detecting rogue updates via differential data-flow analysis
  using trust domains.
\newblock In {\em Proceedings of the IEEE/ACM 46th International Conference on
  Software Engineering}, pages 1--13, 2024.

\bibitem{stefan2014protecting}
Deian Stefan, Edward~Z Yang, Petr Marchenko, Alejandro Russo, Dave Herman, Brad
  Karp, and David Mazieres.
\newblock Protecting users by confining {JavaScript} with {COWL}.
\newblock In {\em 11th USENIX Symposium on Operating Systems Design and
  Implementation (OSDI 14)}, pages 131--146, 2014.

\bibitem{pmforce}
Marius Steffens and Ben Stock.
\newblock Pmforce: Systematically analyzing postmessage handlers at scale.
\newblock In {\em Proceedings of the 2020 ACM SIGSAC Conference on Computer and
  Communications Security}, CCS '20, page 493–505, 2020.

\bibitem{cncfcatalog}
{CNCF}~Security {TAG}.
\newblock Catalog of supply chain compromises.
\newblock
  \url{https://github.com/cncf/tag-security/blob/06147d5/supply-chain-security/compromises}.

\bibitem{terrace2012javascript}
Jeff Terrace, Stephen~R Beard, and Naga Praveen~Kumar Katta.
\newblock {JavaScript} in {JavaScript} (js.js): Sandboxing third-party scripts.
\newblock In {\em 3rd USENIX Conference on Web Application Development (WebApps
  12)}, pages 95--100, 2012.

\bibitem{vasilakis2018breakapp}
Nikos Vasilakis, Ben Karel, Nick Roessler, Nathan Dautenhahn, Andr{\'e} DeHon,
  and Jonathan~M Smith.
\newblock {BreakApp}: Automated, flexible application compartmentalization.
\newblock In {\em NDSS}, 2018.

\bibitem{vasilakis2021preventing}
Nikos Vasilakis, Cristian-Alexandru Staicu, Grigoris Ntousakis, Konstantinos
  Kallas, Ben Karel, Andr{\'e} DeHon, and Michael Pradel.
\newblock Preventing dynamic library compromise on node. js via rwx-based
  privilege reduction.
\newblock In {\em Proceedings of the 2021 ACM SIGSAC Conference on Computer and
  Communications Security}, pages 1821--1838, 2021.

\bibitem{wang2023hodor}
Wenya Wang, Xingwei Lin, Jingyi Wang, Wang Gao, Dawu Gu, Wei Lv, and Jiashui
  Wang.
\newblock Hodor: Shrinking attack surface on node. js via system call
  limitation.
\newblock In {\em Proceedings of the 2023 ACM SIGSAC Conference on Computer and
  Communications Security}, pages 2800--2814, 2023.

\bibitem{openssfcasestudies}
{OpenSSF} Package~Analysis {WG}.
\newblock {OpenSSF} package analysis case studies.
\newblock
  \url{https://github.com/ossf/package-analysis/blob/c4af43d/docs/case_studies.md}.

\bibitem{wyss2022wolf}
Elizabeth Wyss, Alexander Wittman, Drew Davidson, and Lorenzo De~Carli.
\newblock Wolf at the door: Preventing install-time attacks in npm with
  {Latch}.
\newblock In {\em Proceedings of the 2022 ACM on Asia Conference on Computer
  and Communications Security}, pages 1139--1153, 2022.

\bibitem{Zimmermann19}
Markus Zimmermann, Cristian-Alexandru Staicu, Cam Tenny, and Michael Pradel.
\newblock Small world with high risks: A study of security threats in the npm
  ecosystem.
\newblock In {\em 28th USENIX Security Symposium (USENIX Security 19)}, pages
  995--1010. USENIX Association, 2019.

\end{thebibliography}

\end{document}